\documentclass{article}

\usepackage{PRIMEarxiv}

\usepackage[utf8]{inputenc} 
\usepackage[T1]{fontenc}    
\usepackage{hyperref}       
\usepackage{url}            
\usepackage{booktabs}       
\usepackage{amsfonts}       
\usepackage{nicefrac}       
\usepackage{microtype}      
\usepackage{fancyhdr}       
\usepackage{graphicx}       
\graphicspath{{media/}}     
\usepackage[skip=1.5\baselineskip]{caption}
\usepackage[linesnumbered,ruled,vlined]{algorithm2e}	
\usepackage{soul}
\SetKwInput{KwInput}{Input}                
\SetKwInput{KwOutput}{Output}              
\usepackage{multirow}
\usepackage{xcolor}
\usepackage{amsmath}
\usepackage{subcaption}
\usepackage{cite}

\title{Brezinski Inverse and Geometric Product-Based Steffensen's Methods for Image Reverse Filtering
}

\author{
  Guang~Deng \\
  Department of Engineering, La Trobe University \\
  Bundoora, VIC 3086, Australia\\
  \texttt{d.deng@latrobe.edu.aul} \\
}

\begin{document}
\maketitle

\begin{abstract}
This work develops extensions of Steffensen's method to provide new tools for solving the semi-blind image reverse filtering problem.  Two extensions are presented: a parametric Steffensen's method for accelerating the Mann iteration, and a family of 12 Steffensen's methods for vector variables. The development is based on Brezinski inverse and geometric product vector inverse. Variants of these methods are presented with adaptive parameter setting and first-order method acceleration. Implementation details, complexity, and convergence are discussed, and the proposed methods are shown to generalize existing algorithms. A comprehensive study of 108 variants of the vector Steffensen's methods is presented in the Supplementary Material. Representative results and comparison with current state-of-the-art methods demonstrate that the vector Steffensen's methods are efficient and effective tools in reversing the effects of commonly used filters in image processing. 
\end{abstract}


\section{Introduction}

Solving nonlinear systems of equations is essential for addressing
numerous scientific and engineering problems. For instance, one significant
challenge is to estimate the image $\boldsymbol{x}$ in a semi-blind
image reverse filtering problem, where we have an unknown but available
filter $f$ and the observation $\boldsymbol{x}_{0}=f(\boldsymbol{x})$.
This problem can be viewed as solving a nonlinear system of equations.

While numerical methods \cite{ostrowski1966} for solving nonlinear
systems of equations usually involve iterations, their speed of convergence
can be accelerated by many well-established methods such as Aitken
\cite{aitken1925bernoulli}, Steffensen \cite{steffensen1933remarks},
Anderson \cite{anderson1965iterative} and Wynn \cite{wynn1962acceleration}.
The main idea is the transformation/extrapolation of a sequence to
achieve acceleration \cite{SIAM-100-Digit,Brezinski2000}. These convergence
acceleration techniques have also been applied to machine learning
and signal processing domains \cite{optimizationAlgo, Zhang2020GloballyCT,Chen2021RobustSG,Cheby-IEEE-SP-Lett,Anderson_cheby },
to name a few.

In this work, we tackle the problem of extending Steffensen's method
from its original scalar variable version to a vector variable version,
with a specific application in solving a semi-blind image reverse
filtering problem. There are three main motivations and related areas
of research for this work.
\begin{itemize}
\item From a practical application point of view, several iterative methods
\cite{Tmethod,Rmethod,TDAmethod,BELYAEV2022116833,cleanNoiseFilters1,polyBlur}
have been recently published to tackle the problem of semi-blind image
reverse filtering. Some of these methods such as \cite{Tmethod,Rmethod,TDAmethod}
treat every pixel as an independent variable, while other methods
such as the P-method and the S-method \cite{Pmethod} have a parameter,
that is expressed as a ratio of matrix 2-norms. In general, the P-method
and S-method have produced some of the best results for reversing
the effects of a wide range of linear and nonlinear filters. However,
there is no rigorous justification of the definition of the parameter
which is also the computational bottleneck due to $O(n^{2})$ complexity
of calculating the matrix 2-norm. The first motivation is thus to
developing new tools which are efficient and effective for solving
this problem.
\item From an algorithm development point of view, extension of acceleration
methods such as Aitken \cite{aitken1925bernoulli}, Steffensen \cite{steffensen1933remarks}
and Wynn \cite{wynn1961epsilon} from scalar variable to vector variable
has been a continuous research area since the pioneering work of Wynn
\cite{wynn1962acceleration}. The main mathematical tool is the definition
of a vector inverse such as Samelson inverse \cite{wynn1962acceleration,matrixTheory_Gentle}
and Brezinski inverse \cite{Brezinski1977,Sidi2017}. Many variants
of vectorization of these acceleration methods have been developed
and some of them are summarized in references \cite{Macleod1986}
and \cite{iterativeFP_Acc}. The former presented 9 variants of Aitken's
method, including methods of vector variable based on Samelson's inverse,
while the latter presented 7 vector variable Aitken/Steffensen's methods
based on both Samelson's and Brezinski's inverse. Since these methods
have been developed from different perspectives, the second motivation
is thus to conduct a systematic study of extending the widely used
Steffensen's method to vector variables.
\item From developing fast algorithms point of view, there are many well
established techniques for accelerating iterative algorithms \cite{optimizationAlgo},
some of these have been used to accelerate iterative algorithms for
solving the semi-blind image reverse filtering problem \cite{TDAmethod,Deng2023SIVP}.
The third motivation is thus to study further acceleration of the
vector variable Steffensen's method.
\end{itemize}
Main contributions and organization of this paper are listed as follows.
\begin{itemize}
\item In Section \ref{subsec:The-Newton-Aitken-method}, we introduce a
positive parameter to Steffensen's method, which is called the parametric
Steffensen's method. It reduces to the original Steffensen's method
when the parameter is set to 1. We demonstrate that, from the perspective
of solving fixed-point problems, the original Steffensen's method
can be seen as an acceleration of the Picard iteration \cite{kelley1995iterative},
whereas the parametric Steffensen's method is an acceleration of the
Mann iteration \cite{MannIteration}.
\item In Sections \ref{subsec:Using-the-Samelson}, we present a comprehensive
study on the vectorization of  Steffensen's method using the Brezinski
inverse. While previous research has explored the application of this
method, we conduct an exhaustive investigation by exploring all possible
combinations of three versions of Steffensen's method and three versions
of the Brezinski inverse. As a result, we derive a family of 12 methods
(Table \ref{tab:A-family-of}), comprising 7 previously published
methods and 5 new methods, offering a significant contribution to
the field.
\item Section \ref{subsec:Using-the-notion} employs the concept of geometric
product \cite{Gull1993-GULINA} as a means to define the vector inverse
for the vectorization of Steffensen's methods. As a result, we obtain
3 new variants of Steffensen's method. To the best of our knowledge,
this is a new application of geometric product in this area of research.
\item Section \ref{subsec:Generalized-Mann-iteration} provides a fresh
perspective on the vector variable Steffensen's methods by demonstrating
how they generalize Mann iteration \cite{MannIteration} and relate
to nonlinear extrapolation techniques such as Anderson \cite{anderson1965iterative}
and Wynn \cite{wynn1962acceleration}.
\item Section \ref{subsec:Using-Nesterov-acceleration} presents a further
acceleration of the proposed iteration methods by two classes of techniques
that can potentially enhance the convergence speed of vector variable
Steffensen methods in image reverse filtering. The first class includes
two techniques, namely exponential decay and Chebyshev sequence \cite{Cheby-IEEE-SP-Lett,Deng2023SIVP}.
They take advantage of the built in parameter and adjust it during
each iteration. In contrast, the second class is the first-order method
\cite{AFM2018}. It includes the Nesterov method \cite{Nesterov1983AMF,FISTA}
as a special case. It is parameter-free, offering distinct advantages.
\item Section \ref{sec:Application-to-image} presents applications of the
developed iteration schemes to solve the semi-blind image inverse
filtering problem. We show details of implementation, discuss computation
complexity and convergence, and show that some of the iteration schemes
are generalization of existing iterative algorithms. We also highlight
the challenges in using quasi-Newton's methods like Broyden's method
to address this problem, further emphasizing the significance of our
proposed methods.
\item Section \ref{sec:Results} and the accompanying Supplementary Material
present representative results and comparison with current state-of-the-art
methods. These results demonstrate that the vector variable Steffensen's
methods are effective and efficient tools in reversing the effects
of four commonly used filters in image processing.
\end{itemize}
We adopt the following notations. A bold-face character such as $\boldsymbol{x}$
represents a vector. The transpose is $\boldsymbol{x}^{T},$ the inner
product between two vectors is $\boldsymbol{x}^{T}\boldsymbol{y}$,
and the squared $l_{2}$-norm is defined as $||\boldsymbol{x}_{n}||^{2}=\boldsymbol{x}_{n}^{T}\boldsymbol{x}_{n}$,
where the subscript $n$ is used to represent the iteration index.
Scalars are represented as non-bold face characters. 

When Steffensen's method is used for accelerating fixed-point iteration,
it has the same form as that of Aitken's method, although it is originally
developed to accelerate an iterative root finding method. A historic
account can be found in Ostrowski's classic textbook (Appendix E)
\cite{ostrowski1966}. In literature, the vector variable methods
are referred to Aitken in \cite{Macleod1986} and a combination of
Aitken and Steffensen \cite{iterativeFP_Acc}. In this work, we will
use Steffensen to refer to the developed method to highlight the problem
of semi-blind image reverse filtering being formulated as solving
a system of nonlinear equations.

\section{Main results\label{sec:Main-results}}

\subsection{The parametric Steffensen's method\label{subsec:The-Newton-Aitken-method} }

There are two equivalent versions of Steffensen's method. For accelerating
a fixed-point iteration $x_{n+1}=\phi(x_{n})$ called Picard iteration,
Steffensen's method is given by 
\begin{equation}
x_{n+1}=x_{n}-\frac{(\phi(x_{n})-x_{n})^{2}}{\phi(\phi(x_{n}))-2\phi(x_{n})+x_{n}}.\label{eq:a1}
\end{equation}
which is also known as Aitken's $\Delta^{2}$ method \cite{aitken1925bernoulli}.
For solving a root finding problem $g(x)=0$, Steffensen's method
can be motivated from Newton's method
\begin{equation}
x_{n+1}=x_{n}-\frac{g(x_{n})}{g'(x_{n})},\label{eq:newton}
\end{equation}
where the gradient $g'(x_{n})$ is approximated by
\begin{equation}
g'(x_{n})=\frac{g(x_{n}+g(x_{n}))-g(x_{n})}{g(x_{n})}.\label{eq:steff_1}
\end{equation}
The root finding problem is related to the fixed-point problem by
writing 
\begin{equation}
g(x)=\phi(x)-x\label{eq:root_finding-fixed-point}
\end{equation}
such that the fixed-point $x^{*}=\phi(x^{*})$ is also the root $g(x^{*})=0$.
Substitution of (\ref{eq:root_finding-fixed-point}) into (\ref{eq:newton})
and (\ref{eq:steff_1}), we can derive (\ref{eq:a1}). We remark that
the connection between root finding and fixed-point iteration presented
here follows the classic textbook \cite{ostrowski1966}. The underlying
assumption is that the iteration $x_{n+1}=\phi(x_{n})$ converges
to a fixed point.

We now generalize Steffensen's method by introducing a positive parameter
$\mu$ in the approximation of $g'(x)$ stated in (\ref{eq:steff_1})
as follows.
\begin{equation}
g'(x_{n})=\frac{g(x_{n}+\mu g(x_{n}))-g(x_{n})}{\mu g(x_{n})}\label{eq:mu}
\end{equation}
We define a new function
\begin{equation}
\varphi(x)=x+\mu g(x)=x+\mu(\phi(x)-x).\label{eq:mann0-1}
\end{equation}
Using $\varphi(x)$, we can rewrite (\ref{eq:mu}) as the following
\begin{equation}
g'(x_{n})=\frac{\varphi(\varphi(x_{n}))-2\varphi(x_{n})+x_{n}}{\mu(\varphi(x_{n})-x_{n})},\label{eq:ps0}
\end{equation}
Using (\ref{eq:root_finding-fixed-point}) and (\ref{eq:mann0-1}),
we can also write 
\begin{equation}
g(x_{n})=\frac{1}{\mu}(\varphi(x_{n})-x_{n}).\label{eq:gx}
\end{equation}
Substitution of (\ref{eq:ps0}) and (\ref{eq:gx}) into (\ref{eq:newton}),
we have following the iteration:
\begin{equation}
x_{n+1}=x_{n}-\frac{(\varphi(x_{n})-x_{n})^{2}}{\varphi(\varphi(x_{n}))-2\varphi(x_{n})+x_{n}}.\label{eq:ps}
\end{equation}
which is called parametric Steffensen's method.

Comparing (\ref{eq:a1}) with (\ref{eq:ps}), we can see that the
latter is Steffensen's acceleration of the fixed-point iteration $x_{n+1}=\varphi(x_{n})$
which is called Mann iteration \cite{MannIteration,Borwein1991}.
Mann iteration includes Picard iteration \cite{kelley1995iterative}
as a special case when $\mu=1$. In addition, for solving a fixed-point
$x=\phi(x)$, we can use either the Picard iteration with function
$\phi(x)$ or the Mann iteration with function $\varphi(x)$. The
original Steffensen's method is developed to accelerate the Picard
iteration. The parametric Steffensen's method is developed to accelerate
of Mann iteration with function $\varphi(x)$. Since both the original
and parametric Steffensen's method are of the same form, we only discuss
the vectorization of the original Steffensen's method in next section.
We discuss the parametric version in Section \ref{subsec:Using-Nesterov-acceleration}.

\subsection{Vectorization\label{subsec:Vetorization}}

There are three equivalent versions of Steffensen's method:
\begin{itemize}
\item A1: $x_{n+1}=x_{n}+\omega_{n}$, where $\omega_{n}=a^{2}c^{-1}$.
\item A2: $x_{n+1}=\phi(x_{n})+\omega_{n}$, where $\omega_{n}=abc^{-1}$.
\item A3: $x_{n+1}=\phi(\phi(x_{n}))+\omega_{n}$, where $\omega_{n}=b^{2}c^{-1}$.
\end{itemize}
For scalar variable $\phi:R\rightarrow R$ and for vector variable
$\phi:R^{N}\rightarrow R^{N}$, scalar and vector variables are defined
as
\begin{center}
\begin{tabular}{c|c}
\hline 
$a=\phi(x)-x$ & $\boldsymbol{a}=\phi(\boldsymbol{x})-\boldsymbol{x}$\tabularnewline
\hline 
$b=\phi(\phi(x))-\phi(x)$ & $\boldsymbol{b}=\phi(\phi(\boldsymbol{x}))-\phi(\boldsymbol{x})$\tabularnewline
\hline 
$c=a-b$ & $\boldsymbol{c}=\boldsymbol{a}-\boldsymbol{b}$\tabularnewline
\hline 
\end{tabular}
\par\end{center}

The first column of Table \ref{tab:A-summary-of-2} shows different
ways to represent $\omega_{n}$ due to the three versions of Steffensen's
method. In fact, method A2 has three more variants A2.4: $bac^{-1}$,
A2.5: $c^{-1}ab$ and A2.6: $c^{-1}ba$. They are not shown in the
Table because when using the Brezinski inverse, they are equivalent
to the cases A2.1-A2.3. However, all cases including (A2.4-A2.6) are
used in the development based on the geometric product.

The generalization of Steffensen's method to handle vector variables
requires two key steps.
\begin{enumerate}
\item Replacing all scalar variables with the corresponding vector variables.
\item Dealing with the inverse of the vector $\boldsymbol{c}^{-1}$.
\end{enumerate}
For example, to generalize the scalar $\omega_{n}$ of the A2.1 method
to the vector variable $\boldsymbol{\omega}_{n}$, we must first convert
the three scalars ($a,b,c)$ to vectors ($\boldsymbol{a}$, $\boldsymbol{b}$,
$\boldsymbol{c}$) and write $\boldsymbol{\omega}_{n}=\boldsymbol{a}\boldsymbol{b}\boldsymbol{c}^{-1}$.
An essential requirement is that the multiplication of the first two
vector variables must be a scalar, such that $\boldsymbol{\omega}_{n}=(\boldsymbol{a}^{T}\boldsymbol{b})\boldsymbol{c}^{-1}$.
We then use two approaches to deal with the vector inverse $\boldsymbol{c}^{-1}$:
Brezinski inverse (section \ref{subsec:Using-the-Samelson}) and geometric
product (\ref{subsec:Using-the-notion}).

\subsubsection{Using Brezinski inverse\label{subsec:Using-the-Samelson}}

Brezinski inverse \cite{Brezinski1977,Smith1987} is defined for a
pair of vectors $(\boldsymbol{x},\boldsymbol{y})$ ( $\boldsymbol{x}^{T}\boldsymbol{y}\ne0$)
as another pair of vectors $(\boldsymbol{y}^{-1},\boldsymbol{x}^{-1})$
which are defined as
\begin{equation}
\boldsymbol{x}^{-1}=\frac{\boldsymbol{y}}{\boldsymbol{y}^{T}\boldsymbol{x}},\label{eq:samelson1}
\end{equation}
and
\begin{equation}
\boldsymbol{y}^{-1}=\frac{\boldsymbol{x}}{\boldsymbol{x}^{T}\boldsymbol{y}}.\label{eq:samelson1-1}
\end{equation}
They are the inverse vectors with respect to $\boldsymbol{y}$ and
$\boldsymbol{x}$, respectively. The Samelson inverse \cite{wynn1962acceleration,Sidi2017,matrixTheory_Gentle},
can be regarded as a special case when $\boldsymbol{x}=\boldsymbol{y}$
such that 
\begin{equation}
\boldsymbol{x}^{-1}=\frac{\boldsymbol{x}}{\boldsymbol{x}^{T}\boldsymbol{x}}\label{eq:samelson1-2}
\end{equation}
Brezinski inverse is abbreviated as B-inverse in this paper.

We now apply the B-inverse to define $\boldsymbol{c}^{-1}$. In this
study, we explore three variants of the B-inverse, labelled as B1,
B2, and B3. They are listed in Table \ref{tab:A-summary-of-2}, which
also presents the results of resultant the vector $\boldsymbol{\omega}_{n}$
for all combinations of Steffensen's methods and B-inverses.\renewcommand{\arraystretch}{2.5}

\begin{table} 
\begin{centering} 
\begin{tabular}{|c|c||c|c|c|} 
\hline  
\multirow{3}{*}{Cases} 
& \multirow{3}{*}{$\omega_{n}$} 
& \multicolumn{3}{c|}{$\boldsymbol{\omega}_{n}$ using the B-inverse}
\tabularnewline \cline{3-5} \cline{4-5} \cline{5-5}   &  & B1 & B2 & B3\tabularnewline 
\cline{3-5} \cline{4-5} \cline{5-5}   &  
& $\displaystyle \boldsymbol{c}^{-1}=\frac{\boldsymbol{c}}{||\boldsymbol{c}||^{2}}$ 
& $\displaystyle \boldsymbol{c}^{-1}=\frac{\boldsymbol{a}}{\boldsymbol{a}^{T}\boldsymbol{c}}$ 
& $\displaystyle \boldsymbol{c}^{-1}=\frac{\boldsymbol{b}}{\boldsymbol{b}^{T}\boldsymbol{c}}$
\tabularnewline \hline  \hline  
A1.1 & $aac^{-1}$ & $\displaystyle \frac{||\boldsymbol{a}||^{2}}{||\boldsymbol{c}||^{2}}\boldsymbol{c}$ 
& \textcolor{brown}{$\displaystyle  \frac{||\boldsymbol{a}||^{2}}{\boldsymbol{a}{}^{T}\boldsymbol{c}}\boldsymbol{a}$} 
& \textcolor{blue}{$\displaystyle \frac{||\boldsymbol{a}||^{2}}{\boldsymbol{b}^{T}\boldsymbol{c}}\boldsymbol{b}$}
\tabularnewline \hline  
\textcolor{black}{A1.2} & $ac^{-1}a$ 
& \textcolor{brown}{$\displaystyle \frac{\boldsymbol{a}^{T}\boldsymbol{c}}{||\boldsymbol{c}||^{2}}\boldsymbol{a}$} 
& \textcolor{brown}{$\displaystyle \frac{||\boldsymbol{a}||^{2}}{\boldsymbol{a}{}^{T}\boldsymbol{c}}\boldsymbol{a}$} 
& \textcolor{brown}{$\displaystyle \frac{\boldsymbol{a}^{T}\boldsymbol{b}}{\boldsymbol{b}^{T}\boldsymbol{c}}\boldsymbol{a}$}\tabularnewline \hline  
A2.1 & $abc^{-1}$ 
& $\displaystyle \frac{\boldsymbol{a}^{T}\boldsymbol{b}}{||\boldsymbol{c}||^{2}}\boldsymbol{c}$ 
& \textcolor{brown}{$\displaystyle \frac{\boldsymbol{a}^{T}\boldsymbol{b}}{\boldsymbol{a}{}^{T}\boldsymbol{c}}\boldsymbol{a}$} 
& \textcolor{blue}{$\displaystyle \frac{\boldsymbol{a}^{T}\boldsymbol{b}}{\boldsymbol{b}^{T}\boldsymbol{c}}\boldsymbol{b}$}
\tabularnewline \hline  
A2.2 & $ac^{-1}b$ 
& \textcolor{blue}{$\displaystyle \frac{\boldsymbol{a}^{T}\boldsymbol{c}}{||\boldsymbol{c}||^{2}}\boldsymbol{b}$} 
& \textcolor{blue}{$\displaystyle \frac{||\boldsymbol{a}||^{2}}{\boldsymbol{a}{}^{T}\boldsymbol{c}}\boldsymbol{b}$} 
& \textcolor{blue}{$\displaystyle \frac{\boldsymbol{a}^{T}\boldsymbol{b}}{\boldsymbol{b}^{T}\boldsymbol{c}}\boldsymbol{b}$}
\tabularnewline \hline  
\textcolor{black}{A2.3} & $bc^{-1}a$ 
& \textcolor{brown}{$\displaystyle\frac{\boldsymbol{b}^{T}\boldsymbol{c}}{||\boldsymbol{c}||^{2}}\boldsymbol{a}$} 
& \textcolor{brown}{$\displaystyle\frac{\boldsymbol{a}^{T}\boldsymbol{b}}{\boldsymbol{a}{}^{T}\boldsymbol{c}}\boldsymbol{a}$} 
& \textcolor{brown}{$\displaystyle\frac{||\boldsymbol{b}||^{2}}{\boldsymbol{b}^{T}\boldsymbol{c}}\boldsymbol{a}$}
\tabularnewline \hline  
A3.1 & $bbc^{-1}$ 
& $\displaystyle \frac{||\boldsymbol{b}||^{2}}{||\boldsymbol{c}||^{2}}\boldsymbol{c}$ 
& \textcolor{brown}{$\displaystyle \frac{||\boldsymbol{b}||^{2}}{\boldsymbol{a}{}^{T}\boldsymbol{c}}\boldsymbol{a}$} 
& \textcolor{blue}{$\displaystyle\frac{||\boldsymbol{b}||^{2}}{\boldsymbol{b}^{T}\boldsymbol{c}}\boldsymbol{b}$}
\tabularnewline \hline  
A3.2 & $bc^{-1}b$ 
& \textcolor{blue}{$\displaystyle\frac{\boldsymbol{b}^{T}\boldsymbol{c}}{||\boldsymbol{c}||^{2}}\boldsymbol{b}$} 
& \textcolor{blue}{$\displaystyle\frac{\boldsymbol{a}^{T}\boldsymbol{b}}{\boldsymbol{a}{}^{T}\boldsymbol{c}}\boldsymbol{b}$} 
& \textcolor{blue}{ \textcolor{blue}{$\displaystyle \frac{||\boldsymbol{b}||^{2}}{\boldsymbol{b}^{T}\boldsymbol{c}}\boldsymbol{b}$}}\tabularnewline \hline  
\end{tabular} 
\par\end{centering} 
\bigskip
\caption{\label{tab:A-summary-of-2}A summary of converting the scalar variable $\omega_{n}$ to the vector variable $\boldsymbol{\omega}_{n}$ for all combinations of Steffensen's cases and the Brezinski-inverses. The vector $\boldsymbol{\omega}_{n}$ has a unified form of $\boldsymbol{\omega}_{n}=\lambda_n \boldsymbol{z}$,$\boldsymbol{z}$ can be  \textcolor{brown}{$\boldsymbol{a}$} or \textcolor{blue}{$\boldsymbol{b}$} or \textcolor{black}{$\boldsymbol{c}$}.} 
\end{table}

As an example, the scalar variable case A1.1 is given by 
\begin{equation}
x_{n+1}=x_{n}+a^{2}c^{-1}.\label{eq:a1.1}
\end{equation}
The corresponding vector variable case using inverse B2 $\boldsymbol{c}^{-1}=\boldsymbol{a}/\boldsymbol{a}{}^{T}\boldsymbol{c}$
is given by
\begin{align}
\boldsymbol{x}_{n+1} & =\boldsymbol{x}_{n}+||\boldsymbol{a}||^{2}\boldsymbol{c}^{-1}\nonumber \\
 & =\boldsymbol{x}_{n}+\frac{||\boldsymbol{a}||^{2}}{\boldsymbol{a}{}^{T}\boldsymbol{c}}\boldsymbol{a}.\label{eq:a1.1-s2}
\end{align}
This iteration is denoted A1.1-B2 due to the combination of the two
cases. Similarly, the scalar variable iteration of case A2.1 is given
by 
\begin{equation}
x_{n+1}=\phi(x_{n})+abc^{-1}\label{it2}
\end{equation}
and its corresponding vector variable iteration by using inverse B2
is given by
\begin{align}
\boldsymbol{x}_{n+1} & =\phi(\boldsymbol{x}_{n})+(\boldsymbol{a}^{T}\boldsymbol{b})\boldsymbol{c}^{-1}\nonumber \\
 & =\phi(\boldsymbol{x}_{n})+\frac{\boldsymbol{a}^{T}\boldsymbol{b}}{\boldsymbol{a}{}^{T}\boldsymbol{c}}\boldsymbol{a}.\label{eq:it1}
\end{align}
This iteration is called A2.1-B2. By using $\phi(\boldsymbol{x}_{n})=\boldsymbol{x}_{n}+\boldsymbol{a}$
and $\boldsymbol{a}=\boldsymbol{b}+\boldsymbol{c}$, we can show that
iteration (\ref{eq:it1}) is equivalent to iteration (\ref{eq:a1.1-s2}).
Indeed, we can derive (\ref{eq:a1.1-s2}) from (\ref{eq:it1}) as
follows
\begin{align}
\boldsymbol{x}_{n+1} & =\phi(\boldsymbol{x}_{n})+\frac{\boldsymbol{a}^{T}\boldsymbol{b}}{\boldsymbol{a}{}^{T}\boldsymbol{c}}\boldsymbol{a}\nonumber \\
 & =\boldsymbol{x}_{n}+\boldsymbol{a}+\frac{\boldsymbol{a}^{T}\boldsymbol{b}}{\boldsymbol{a}{}^{T}\boldsymbol{c}}\boldsymbol{a}\nonumber \\
 & =\boldsymbol{x}_{n}+\frac{||\boldsymbol{a}||^{2}}{\boldsymbol{a}{}^{T}\boldsymbol{c}}\boldsymbol{a}\label{eq:drive}
\end{align}

From Table \ref{tab:A-summary-of-2}, there are 21 ways to combine
the 7 Steffensen's methods with the 3 B-inverses. Because some of
them lead to the same iteration (e.g., we have shown cases A1.1-B2
and A2.1-B2 are the same), there are only 11 unique iterations. Results
are summarized in Table \ref{tab:A-family-of} where the first column
shows the cases which result in the same iteration. We can classify
these 11 iterations into three types according to the update variables.
The $\mathcal{A}$-type with update variable $\boldsymbol{a}$ has
4 unique cases: $\mathcal{A}1$ to $\mathcal{A}4$, the $\mathcal{B}$-type
with update variable $\boldsymbol{b}$ has 4 unique cases: $\mathcal{B}1$
to $\mathcal{B}4$, and the $\mathcal{C}$-type with update variable
$\boldsymbol{c}$ has 3 unique cases: $\mathcal{C}1$ to $\mathcal{C}3$.
Among them, five iteration algorithms are new. The last row in the
Table is derived from using the geometric product and is discussed
in the next section.
\renewcommand{\arraystretch}{3}
\begin{table*}
\begin{centering}
\begin{tabular}{|c|c|c|c|c|}
\hline 
Cases & Type & Iteration & $\lambda_n$, $\eta_n$  & Notes\tabularnewline
\hline 
\hline 
A1.1-B2, A1.2-B2, A2.1-B2, A2.3-B2 & $\mathcal{A}1$ & \multirow{3}{*}{$\boldsymbol{x}_{n+1}=\boldsymbol{x}_{n}+\lambda_{n}\boldsymbol{a}$} & $\lambda_n=\displaystyle \frac{||\boldsymbol{a}||^{2}}{\boldsymbol{a}{}^{T}\boldsymbol{c}}$ & New\tabularnewline
\cline{1-2} \cline{2-2} \cline{4-5} \cline{5-5} 
A1.2-B1, A2.3-B1 & $\mathcal{A}2$ &  & $\lambda_n=\displaystyle \frac{\boldsymbol{a}^{T}\boldsymbol{c}}{||\boldsymbol{c}||^{2}}$ & \cite{lemarechal1971methode}\tabularnewline
\cline{1-2} \cline{2-2} \cline{4-5} \cline{5-5} 
A1.2-B3, A2.3-B3 & $\mathcal{A}3$ &  & $\lambda_n=\displaystyle \frac{\boldsymbol{a}^{T}\boldsymbol{b}}{\boldsymbol{b}^{T}\boldsymbol{c}}$ & \cite{sedogbo1997some}\tabularnewline
\hline 
A3.1-B2 & $\mathcal{A}4$ & $\boldsymbol{x}_{n+1}=\phi(\phi(\boldsymbol{x}_{n}))+\lambda_{n}\boldsymbol{a}$ & $\lambda_n=\displaystyle \frac{||\boldsymbol{b}||^{2}}{\boldsymbol{a}{}^{T}\boldsymbol{c}}$ & New\tabularnewline
\hline 
A1.1-B3 & $\mathcal{B}1$ & $\boldsymbol{x}_{n+1}=\boldsymbol{x}_{n}+\lambda_{n}\boldsymbol{b}$ & $\lambda_n=\displaystyle \frac{||\boldsymbol{a}||^{2}}{\boldsymbol{b}{}^{T}\boldsymbol{c}}$ & New\tabularnewline
\hline 
A2.2-B1, A3.2-B1 & $\mathcal{B}2$ & \multirow{3}{*}{$\boldsymbol{x}_{n+1}=\phi(\boldsymbol{x}_{n})+\lambda_{n}\boldsymbol{b}$} & $\lambda_n=\displaystyle \frac{\boldsymbol{a}^{T}\boldsymbol{c}}{||\boldsymbol{c}||^{2}}$ & \cite{Jennings1971}(SDM), \cite{irons1969version}, \cite{anderson1965iterative},
\cite{Macleod1986}(M3)\tabularnewline
\cline{1-2} \cline{2-2} \cline{4-5} \cline{5-5} 
A2.1-B3, A2.2-B3, A3.1-B3, A3.2-B3 & $\mathcal{B}3$ &  & $\lambda_n=\displaystyle \frac{\boldsymbol{a}^{T}\boldsymbol{b}}{\boldsymbol{b}^{T}\boldsymbol{c}}$ & \cite{Zienkiewicz1985AcceleratedO}, \cite{Macleod1986}(M4)\tabularnewline
\cline{1-2} \cline{2-2} \cline{4-5} \cline{5-5} 
A2.2-B2, A3.2-B2 & $\mathcal{B}4$ &  & $\lambda_n=\displaystyle \frac{||\boldsymbol{a}||^{2}}{\boldsymbol{a}{}^{T}\boldsymbol{c}}$ & \cite{Morris-extrapolation-epsilon92}, \cite{Jennings1971}(FDM),
\cite{Macleod1986}(M5)\tabularnewline
\hline 
A1.1-B1 & $\mathcal{C}1$ & $\boldsymbol{x}_{n+1}=\boldsymbol{x}_{n}+\lambda_{n}\boldsymbol{c}$ & $\lambda_n=\displaystyle \frac{||\boldsymbol{a}||^{2}}{||\boldsymbol{c}||^{2}}$ & New\tabularnewline
\hline 
A2.1-B1 & $\mathcal{C}2$ & $\boldsymbol{x}_{n+1}=\phi(\boldsymbol{x}_{n})+\lambda_{n}\boldsymbol{c}$ & $\lambda_n=\displaystyle \frac{\boldsymbol{a}^{T}\boldsymbol{b}}{||\boldsymbol{c}||^{2}}$ & New\tabularnewline
\hline 
A3.1-B1 & $\mathcal{C}3$ & $\boldsymbol{x}_{n+1}=\phi(\phi(\boldsymbol{x}_{n}))+\lambda_{n}\boldsymbol{c}$ & $\lambda_n=\displaystyle \frac{||\boldsymbol{b}||^{2}}{||\boldsymbol{c}||^{2}}$ & \cite{Macleod1986}(M6)\tabularnewline
\hline 
\multirow{2}{*}{A1.2-G, A2.2-G, A2.3-G, A3.2-G} & \multirow{2}{*}{$\epsilon$} & \multirow{2}{*}{$\boldsymbol{x}_{n+1}=\phi(\boldsymbol{x}_{n})+\lambda_{n}\boldsymbol{b}-\eta_{n}\boldsymbol{a}$} & $\lambda_n=\displaystyle \frac{||\boldsymbol{a}||^{2}}{||\boldsymbol{c}||^{2}}$ & \multirow{2}{*}{\cite{wynn1962acceleration}, \cite{Macleod1986}(M2)}\tabularnewline
 &  &  & $\eta_n=\displaystyle \frac{||\boldsymbol{b}||^{2}}{||\boldsymbol{c}||^{2}}$ & \tabularnewline
\hline 
\end{tabular}
\par\end{centering}
\bigskip
\caption{\label{tab:A-family-of}A family of 12 vector variable Steffensen's
methods using Brezinski inverse and geometric product-based method.
The 1st column shows the combinations of Steffensen's method and the
vector inverse. The 2nd column presents the names of each iteration
according to the update variable used. The last column shows where
applicable the reference and the name for the algorithm in the reference,
e.g., {[}24{]}(M5) refers to the 5th method presented in {[}24{]}.}
\end{table*}

\renewcommand{\arraystretch}{1.1}

\subsubsection{Using the notion of geometric product\label{subsec:Using-the-notion}}

The geometric product \cite{Gull1993-GULINA} of two vectors is defined
as
\begin{equation}
\boldsymbol{x}\boldsymbol{y}=\boldsymbol{x}^{T}\boldsymbol{y}+\boldsymbol{x}\wedge\boldsymbol{y}\label{eq:gp1}
\end{equation}
where the second term is called the exterior product and is defined
as 
\begin{equation}
\boldsymbol{x}\wedge\boldsymbol{y}=\frac{1}{2}(\boldsymbol{xy}-\boldsymbol{yx}).\label{eq:extProd}
\end{equation}
Because $\boldsymbol{x}\wedge\boldsymbol{x}=0$, the geometric product
of a vector with itself is the inner product $\boldsymbol{xx=x}^{T}\boldsymbol{x}$.
The geometric product is also associative and distributive in that
$\boldsymbol{x}(\boldsymbol{yz})=(\boldsymbol{xy})\boldsymbol{z}$
and $\boldsymbol{x}(\boldsymbol{y}+\boldsymbol{z})=\boldsymbol{xy}+\boldsymbol{xz}$.

The inverse of a vector in terms of the geometric product is defined
as
\begin{equation}
\boldsymbol{x}^{-1}=\frac{\boldsymbol{x}}{||\boldsymbol{x}||^{2}}\label{eq:gp6}
\end{equation}
such that
\begin{equation}
\boldsymbol{x}^{-1}\boldsymbol{xy}=\frac{1}{||\boldsymbol{x}||^{2}}(\boldsymbol{xx})\boldsymbol{y}=\frac{1}{||\boldsymbol{x}||^{2}}||\boldsymbol{x}||^{2}\boldsymbol{y}=\boldsymbol{y.}\label{eq:gp7}
\end{equation}
It is interesting to know that (\ref{eq:gp6}) is of the same form
as the Samelson inverse. For the vectorization of Steffensen's method,
we also need the following property 

\begin{equation}
\boldsymbol{x}\boldsymbol{y}\boldsymbol{x}=2(\boldsymbol{x}^{T}\boldsymbol{y})\boldsymbol{x}-||\boldsymbol{x}||^{2}\boldsymbol{y.}\label{eq:gp5}
\end{equation}
It can be derived as the following:
\begin{align}
\boldsymbol{x}\boldsymbol{y}\boldsymbol{x} & =\boldsymbol{x}(\boldsymbol{y}\boldsymbol{x})\nonumber \\
 & =\boldsymbol{x}(\boldsymbol{y}^{T}\boldsymbol{x}+\boldsymbol{y}\wedge\boldsymbol{x})\nonumber \\
 & =\boldsymbol{x}\left(\boldsymbol{y}^{T}\boldsymbol{x}+\frac{1}{2}(\boldsymbol{y}\boldsymbol{x}-\boldsymbol{x}\boldsymbol{y})\right)\nonumber \\
 & =(\boldsymbol{x}^{T}\boldsymbol{y})\boldsymbol{x}+\frac{1}{2}\left(\boldsymbol{x}\boldsymbol{y}\boldsymbol{x}-||\boldsymbol{x}||^{2}\boldsymbol{y}\right).\label{eq:gp6-1}
\end{align}
Rearrangement and simplification of (\ref{eq:gp6-1}) yields (\ref{eq:gp5}).

To generalize Steffensen's method to handle vector variables, multiplication
operation is treated as the geometric product. As an example, we derive
the vector variable algorithm for the case A1.2 by first generalizing
the scalar variable to vector variable: $\omega_{n}=ac^{-1}b$ $\rightarrow$
$\boldsymbol{\omega}_{n}=\boldsymbol{a}\boldsymbol{c}^{-1}\boldsymbol{a}$.
Using equations (\ref{eq:gp6}) and (\ref{eq:gp5}), the iteration
is then given by
\begin{equation}
\boldsymbol{x}_{n+1}=\boldsymbol{x}_{n}+\frac{(2\boldsymbol{a}^{T}\boldsymbol{c})\boldsymbol{a-}||\boldsymbol{a}||^{2}\boldsymbol{c}}{||\boldsymbol{c}||^{2}}.\label{eq:up1}
\end{equation}
Using $\boldsymbol{x}_{n}=\phi(\boldsymbol{x}_{n})-\boldsymbol{a}$
and $\boldsymbol{c=a-b}$, we can show that the above iteration can
be written as
\begin{equation}
\boldsymbol{x}_{n+1}=\phi(\boldsymbol{x}_{n})+\frac{||\boldsymbol{a}||^{2}\boldsymbol{b}-||\boldsymbol{b}||^{2}\boldsymbol{a}}{||\boldsymbol{c}||^{2}}.\label{eq:up2}
\end{equation}
Appendix shows the vectorization for all cases (A1 to A3). There are
only three unique iteration algorithms which are shown in Table \ref{tab:Summary.}.
In this Table, we add ``-G'' to the name of the case to make it
different from that using the B-inverse.
\begin{table}
\begin{centering}
\begin{tabular}{|c|c|}
\hline 
Cases & Iteration\tabularnewline
\hline 
\hline 
A1.1-G, A2.1-G, A2.6-G & same as $\mathcal{C}_{1}$\tabularnewline
\hline 
A1.2-G, A2.2-G, A2.3-G, A3.2-G & same as $\epsilon$\tabularnewline
\hline 
A2.4-G, A2.5-G, A3.1-G & same as $\mathcal{C}_{3}$\tabularnewline
\hline 
\end{tabular}
\par\end{centering}
\bigskip
\caption{\label{tab:Summary.}Iteration algorithms derived from all cases of
Steffensen's method using vector inverse defined using the notion
of geometric product. Iteration equations are presented in Table \ref{tab:A-family-of}.}
\end{table}
The first one and the third one have been derived by using the B-inverse,
while the second one is known as Wynn's $\epsilon$-method \cite{wynn1962acceleration}.

\subsubsection{Comment on the two approaches}

When we use Brezinski inverse or the inverse defined for the geometric
product, we assume one of these conditions are satisfied: $||\boldsymbol{c}||^{2}\ne0$,
$\boldsymbol{a}^{T}\boldsymbol{c}\ne0$, and $\boldsymbol{b}^{T}\boldsymbol{c}\ne0$,
depending on the variants. However, in image reverse filtering applications,
this condition may not be satisfied. A hard limiter (see equation
(\ref{eq:hl})) is used to deal with this problem.

Brezinski inverse is a general method for defining the inverse of
a vector $\boldsymbol{c}$ by pairing it with another vector $\boldsymbol{v}$
to form a vector pair $(\boldsymbol{v},\boldsymbol{c})$. In this
work, we have examined the three natural choices $\boldsymbol{v=c}$,
$\boldsymbol{v}=\boldsymbol{a}$, and $\boldsymbol{v}=\boldsymbol{b}.$
The first choice leads to the Samelson inverse which is related to
the Moore-Penrose pseudo inverse \cite{matrixTheory_Gentle}. The
meaning and properties of the other two choices are unclear and need
further investigation. Moreover, the vectorization process using this
method depends on expressing $\boldsymbol{\omega}_{n}$ as a scalar
times a vector. For example, to vectorize $\omega_{n}=a^{2}c^{-1}$,
we can generalize $a^{2}$ as $||\boldsymbol{a}||^{2}$ and write
$\boldsymbol{\omega}_{n}=||\boldsymbol{a}||^{2}\boldsymbol{c}^{-1}$.
On the other hand, the inverse based on the geometric product is the
same as the Samelson inverse and does not require to enforce the scalar-vector
multiplication. All vector operations are performed using the geometric
product.

\subsection{Discussion\label{subsec:Generalized-Mann-iteration}}

We show the connection between the Mann iteration and the vector variable
Steffensen's method. We also show that two well-known extrapolation-based
methods can be derived from the vector variable Steffensen's method.

\subsubsection{Generalized Mann iteration}

For a fixed-point problem $\boldsymbol{x}=\phi(\boldsymbol{x}),$
the iteration function of the Mann iteration $\boldsymbol{x}_{n+1}=M(\boldsymbol{x}_{n})$
is defined as
\begin{equation}
M(\boldsymbol{x};\lambda)=\boldsymbol{x}+\lambda(\phi(\boldsymbol{x})-\boldsymbol{x}),\label{Mann}
\end{equation}
where $\lambda>0$ is a scalar parameter. It can be written in an
equivalent extrapolation form as follows
\begin{equation}
E(\boldsymbol{x},\lambda)=\phi(\boldsymbol{x})+\lambda(\phi(\boldsymbol{x})-\boldsymbol{x}).\label{eq:extrapolation}
\end{equation}
The relationship between the two is:
\begin{equation}
M(\boldsymbol{x},1+\lambda)=E(\boldsymbol{x},\lambda).\label{eq:MannExtrapolation}
\end{equation}
We now discuss iterations presented in Table \ref{tab:A-family-of}.
These iterations are divided into 3 groups according to their relationship
with Mann iteration. 

The first group, which contains iterations $\mathcal{A}1$-$\mathcal{A}3$,
can be regarded as a generalized Mann iteration and can be written
as
\begin{equation}
\boldsymbol{x}_{n+1}=M(\boldsymbol{x}_{n};\lambda_{n}),\label{eq:1st group}
\end{equation}
The generalization is in the sense that equation (\ref{eq:1st group})
has a scalar parameter $\lambda_{n}$ which can be either positive
or negative, depending on the vectors in the current iteration. The
difference of these 3 iterations is the way the scale parameter $\lambda_{n}$
is calculated. 

In addition, it is well known that one of the fundamental idea of
accelerating the convergence of a sequence is through an extrapolation
process \cite{SIAM-100-Digit,Brezinski2000}. For example, Steffensen's
method A2 can be written in a nonlinear extrapolation form as the
following
\begin{equation}
y_{n}=\phi(x_{n})+\lambda_{n}(\phi(\phi(x_{n}))-\phi(x_{n})),\label{eq:ne1}
\end{equation}
where 
\begin{equation}
\lambda_{n}=(\phi(x_{n})-x_{n})/(2\phi(x_{n})-\phi(\phi(x_{n}))-x_{n}).\label{eq:lambda_n}
\end{equation}
The interpretation of the vector variable Steffensen's method as a
generalized Mann iteration not only provides a new connection between
the two methods, but also provides a new insight into the extrapolation
nature of the iteration. 

The second group, which contains iterations $\mathcal{B}2$-$\mathcal{B}4$,
can be regarded as a compound Picard-Mann iteration. It has a Picard-step
followed by a Mann-step 
\begin{equation}
\text{Picard-step}:\quad\boldsymbol{y}_{n}=\phi(\boldsymbol{x}_{n}).\label{eq:Picard}
\end{equation}
\begin{equation}
\text{Mann-step}:\boldsymbol{x}_{n+1}=M(\boldsymbol{y}_{n};\lambda_{n}).\label{eq:Mann1}
\end{equation}
Compound iterations have been studied before. For example, the Ishikawa
iteration \cite{Ishikawa1974} can be written as two Mann steps and
has been extended to many versions \cite{Hassan2020ANF}.
\begin{flushleft}
The third group, which contains iterations $\mathcal{A}4$, $\mathcal{B}1$,
$\mathcal{C}1$-$\mathcal{C}3$, and $\epsilon$, can be written as
follows
\begin{equation}
\mathcal{A}4:\boldsymbol{x}_{n+1}=E(\boldsymbol{x}_{n};\lambda_{n})+\boldsymbol{b}.\label{eq:a4}
\end{equation}
\begin{equation}
\mathcal{B}1:\boldsymbol{x}_{n+1}=M(\boldsymbol{y}_{n},\lambda_{n})-\boldsymbol{a}.\label{eq:b1}
\end{equation}
\begin{equation}
\mathcal{C}1:\boldsymbol{x}_{n+1}=M(\boldsymbol{x}_{n};\lambda_{n})-\lambda_{n}\boldsymbol{b}.\label{g1}
\end{equation}
\begin{equation}
\mathcal{C}2:\boldsymbol{x}_{n+1}=E(\boldsymbol{x}_{n};\lambda_{n})-\lambda_{n}\boldsymbol{b}.\label{eq:g2}
\end{equation}
\begin{equation}
\mathcal{C}3:\boldsymbol{x}_{n+1}=E(\boldsymbol{y}_{n};\theta_{n})-\theta_{n}\boldsymbol{a}\quad(\theta_{n}=-\lambda_{n}).\label{eq:g3}
\end{equation}
\begin{equation}
\epsilon:\boldsymbol{x}_{n+1}=M(\boldsymbol{y}_{n},\lambda_{n})-\eta_{n}\boldsymbol{a}.\label{eq:epsilon}
\end{equation}
They are expressed as either a generalized Mann iteration ($\mathcal{B}1$,
$\mathcal{C}1$ and $\epsilon$) or a nonlinear extrapolation ($\mathcal{A}4$,
$\mathcal{C}2$ and $\mathcal{C}3$) plus a correction term. 
\par\end{flushleft}

\subsubsection{Connections with two nonlinear extrapolation methods\label{subsec:Connections-with-two}}

We briefly discuss two special cases, Wynn's $\epsilon$-method \cite{wynn1962acceleration}
and Anderson's method \cite{anderson1965iterative}. The scalar variable
Wynn's method has the following recursive formula for the extrapolation
\begin{equation}
x_{n,k}=x_{n+1,k-2}+\frac{1}{x_{n+1,k-1}-x_{n,k-1}},\label{eq:wynn}
\end{equation}
where $x_{n,k}$ is the extrapolation of the data point $x_{n}$ by
using $k$ data points $\{x_{m}\}_{m=n:n-k+1}$ with the assumptions
$x_{n,0}=x_{n}$ and $x_{n,-|j|}=0$ for any integer $j$. In terms
of solving the fixed-point problem, we have $x_{n+1}=\phi(x_{n})$.
We can prove that when $k=2$, Wynn's method is 
\begin{equation}
x_{n,2}=x_{n+1}+\frac{(x_{n+2}-x_{n+1})(x_{n+1}-x_{n})}{(2x_{n+1}-x_{n+2}-x_{n})}.\label{eq:wynn-1}
\end{equation}
This is exactly the same as Steffensen's method A2 when we substitute
$x_{n+2}=\phi(\phi(x_{n}))$ and $x_{n+1}=\phi(x_{n})$ into the equation.
It is interesting to note that in Wynn's original work, only the vector
variable $\epsilon$ algorithm (last row in Table \ref{tab:A-family-of})
was derived. In this work, we have derived 4 iterations ($\mathcal{B}2-\mathcal{B}4$,
$\epsilon$). 

Anderson's method for accelerating a fixed-point iteration $\boldsymbol{x}_{n+1}=\phi(\boldsymbol{x}_{n})$
uses the following linear combination to perform an estimate based
on the data set $\{\boldsymbol{x}_{n+k+1}\}$ with $k=0:N-1$
\begin{equation}
\boldsymbol{v}_{n+1}=\sum_{k=0}^{N-1}\theta_{k}\boldsymbol{x}_{n+k+1}.\label{eq:AA1}
\end{equation}
The coefficients $\{\theta_{k}\}$ are determined by solving a minimization
problem
\begin{equation}
\min_{\{\theta_{k}\}}||\sum_{k=0}^{N-1}\theta_{k}(\phi(\boldsymbol{x}_{n+k})-\boldsymbol{x}_{n+k})||^{2}\label{eq:AA2}
\end{equation}
subject to $\sum_{k=0}^{N-1}\theta_{k}=1$. For the case $N=2$ and
using the notation $\boldsymbol{a}=\phi(\boldsymbol{x}_{n})-\boldsymbol{x}_{n}$,
$\boldsymbol{b}=\phi(\boldsymbol{x}_{n+1})-\boldsymbol{x}_{n+1}=\phi(\phi(\boldsymbol{x}_{n}))-\phi(\boldsymbol{x}_{n})$
and $\boldsymbol{c}=\boldsymbol{a-b}$, we can derive
\begin{equation}
\theta_{0}=-\boldsymbol{b}^{T}\boldsymbol{c}/||\boldsymbol{c}||^{2},\label{eq:AA3}
\end{equation}
and
\begin{equation}
\boldsymbol{v}_{n+1}=\theta_{0}\phi(\boldsymbol{x}_{n})+(1-\theta_{0})\phi(\phi(\boldsymbol{x}_{n}))\label{eq:AA4}
\end{equation}
which is the same as the vector variable Steffensen's method $\mathcal{B}2$.

Therefore,  Steffensen's method include some special cases of well
known extrapolation-based acceleration methods such as Wynn's method
and Anderson's method. This work shows such connections. For example,
although  Steffensen's method and Anderson's method are developed
from different considerations, but they share the same form. This
work also provides three more iteration algorithms than the original
Wynn's $\epsilon$-method, leading to an expansion of the toolbox
in accelerating the convergence of sequences. 

\subsection{Further acceleration \label{sec:Acceleration}}

We study two classes of techniques that can potentially enhance the
convergence speed of vector variable Steffensen's methods in image
reverse filtering. The first class includes two techniques, namely
exponential decay and Chebyshev sequence \cite{Cheby-IEEE-SP-Lett,Anderson_cheby},
which take advantage of the built in parameter $\mu$ in the parametric
Steffensen's method and adjust it during each iteration. In contrast,
the second class comprises first-order methods \cite{AFM2018} which
are parameter-free.

\subsubsection{Exponential decay and Chebyshev sequence\label{subsec:Exponential-decay-and}}

\subsubsection*{Exponential decay}

Motivated by the original Mann-iteration \cite{MannIteration} which
sets $\mu_{n}=1/(n+1)$ as a decreasing function of the iteration
index $n$, we study two exponential decay methods to adaptively set
the parameter $\mu$. They are abbreviated as ``ed-1'' and ``ed-2''
which are defined as the following
\begin{equation}
\text{{ed-1:}\ensuremath{\quad}}\mu_{n}=1+\exp(-((2n)/N)^{2})\label{ed-1}
\end{equation}
and
\begin{equation}
\text{{ed-2:}\ensuremath{\quad}}\mu_{n}=2\exp(-((2n)/N)^{2})\label{eq:ed-2}
\end{equation}
where $N$ is a user defined maximum of number of iterations. The
difference between the two is that while ``ed-1'' decreases from
2 to 1, ``ed-2'' decrease from 2 to a small number close to 0. Compared
to the original Mann iteration for which the parameter satisfies $\mu_{n}\le1$,
the parameter for both ``ed-1'' and ``ed-2'' starts from 2 and
drops to 1 at certain point of the iteration. Experimental results
show that the period of iteration with $\mu_{n}>1$ speeds up the
improvement of PSNR.

\subsubsection*{Chebyshev sequence}

In a recent paper \cite{Deng2023SIVP}, we have experimented with
a modified Chebyshev sequence \cite{Cheby-IEEE-SP-Lett} for the acceleration
of fixed-point iteration. In this work, we use the modified Chebyshev
sequence to define $\mu_{n}$ as follows
\begin{equation}
\mu_{n}=2\min\left(1,\frac{1}{1+\cos(2(n+1)\pi/P)}\right)\label{eq:cheby}
\end{equation}
where $P$ is the period of the sequence and is set to $64$ in all
experiments. A detailed discussion of the modified Chebyshev sequence
is presented in \cite{Deng2023SIVP}.

\subsubsection{Accelerated first order method (AFM) \label{subsec:Using-Nesterov-acceleration}}

The vector variable Steffensen's method presented in Table \ref{tab:A-family-of}
can be expressed in the following form
\begin{equation}
\boldsymbol{x}_{n+1}=h(\boldsymbol{x}_{n})+\lambda_{n}d(\boldsymbol{x}_{n}).\label{eq:n1}
\end{equation}
For example, for method $\mathcal{A}1$, we have $h(\boldsymbol{x}_{n})=\boldsymbol{x}_{n}$,
$\lambda_{n}=||\boldsymbol{a}_{n}||^{2}/\boldsymbol{a}_{n}{}^{T}\boldsymbol{c}_{n}$
and $d(\boldsymbol{x}_{n})=\boldsymbol{a}_{n}$. The AFM \cite{AFM2018}
accelerates the above iteration in two steps
\begin{equation}
\boldsymbol{u}_{n+1}=h(\boldsymbol{x}_{n})+\lambda_{n}d(\boldsymbol{x}_{n}),\label{eq:n2}
\end{equation}
and
\begin{equation}
\boldsymbol{x}_{n+1}=\boldsymbol{u}_{n+1}+\beta_{n}(\boldsymbol{u}_{n+1}-\boldsymbol{u}_{n})+\gamma_{n}(\boldsymbol{u}_{n+1}-\boldsymbol{x}_{n}),\label{eq:n3}
\end{equation}
The parameters are updated by
\begin{equation}
\beta_{n}=(t_{n}-1)/t_{n+1},\label{eq:t2}
\end{equation}
 and 
\begin{equation}
\gamma_{n}=t_{n}/t_{n+1},\label{eq:t3}
\end{equation}
where $t_{0}=1$ and $t_{n+1}=\frac{1}{2}(1+\sqrt{1+4t_{n}^{2}})$.

Since $\lim_{n\rightarrow\infty}\beta_{n}=1$ and $\lim_{n\rightarrow\infty}\gamma_{n}=1$,
when $n$ is sufficiently large, an approximation of the AFM is 
\begin{equation}
\boldsymbol{x}_{n+1}\approx3\boldsymbol{u}_{n+1}-\boldsymbol{u}_{n}-\boldsymbol{x}_{n}.\label{eq:n4}
\end{equation}
The Nesterov acceleration \cite{Nesterov1983AMF,FISTA} is a special
case $\gamma_{n}=0$:
\begin{equation}
\boldsymbol{x}_{n+1}=\boldsymbol{u}_{n+1}+\beta_{n}(\boldsymbol{u}_{n+1}-\boldsymbol{u}_{n}),\label{eq:n5}
\end{equation}
and when $n$ is large, we have 
\begin{equation}
\boldsymbol{x}_{n+1}\approx2\boldsymbol{u}_{n+1}-\boldsymbol{u}_{n}.\label{eq:n6}
\end{equation}

\section{Application to image reverse filtering\label{sec:Application-to-image}}

The problem of semi-blind image reverse filtering can be formulated
as solving a system of nonlinear equations: $g(\boldsymbol{x})=\boldsymbol{x}_{0}-f(\boldsymbol{x})=\boldsymbol{0}$,
where $\boldsymbol{x}_{0}$ is the observation and $f$ is the unknown
but available filter function. We can use the methods developed in
the previous section to solve this problem by defining $g(\boldsymbol{x})=\phi(\boldsymbol{x})-\boldsymbol{x}$
such that iteration function is
\begin{equation}
\phi(\boldsymbol{x})=\boldsymbol{x}+\boldsymbol{x}_{0}-f(\boldsymbol{x})\label{eq:phi1}
\end{equation}
and the solution is a Picard iteration $\boldsymbol{x}_{n+1}=\phi(\boldsymbol{x}_{n})$.
We can also use the Mann iteration to solve the same problem by defining
the following iteration function 
\begin{align}
\varphi(\boldsymbol{x}) & =\boldsymbol{x}+\mu(\phi(\boldsymbol{x})-\boldsymbol{x})\nonumber \\
 & =\boldsymbol{x}+\mu(\boldsymbol{x}_{0}-f(\boldsymbol{x})).\label{eq:mann51}
\end{align}
For both iterations, we have to assume for the unknown filter $f$,
the fixed-point $\boldsymbol{x}^{*}$ exists leading to $g(\boldsymbol{x}^{*})=0$
and $\boldsymbol{x}_{0}=f(\boldsymbol{x}^{*})$. 

\subsection{Implementation, complexity and convergence\label{subsec:Implementations-and-complexity}}

\subsubsection{Implementation}

We can apply the vector variable Steffensen's acceleration to both
iteration schemes. Referring to Section \ref{subsec:Vetorization},
we define the 3 variables $\boldsymbol{a}$,\textbf{ $\boldsymbol{b}$
}and $\boldsymbol{c}$ in terms of the filtering problem in Table
\ref{tab:Vector-variables-for}. 
\begin{table}
\begin{centering}
\begin{tabular}{|c|c|}
\hline 
Parameter-free & Parametric\tabularnewline
\hline 
\hline 
$\boldsymbol{a}_{n}=\boldsymbol{x}_{0}-f(\boldsymbol{x}_{n})$ & $\bar{\boldsymbol{a}}_{n}=\mu_{n}\boldsymbol{a}_{n}$\tabularnewline
\hline 
$\boldsymbol{b}_{n}=\boldsymbol{x}_{0}-f(\boldsymbol{x}_{n}+\boldsymbol{a}_{n})$ & $\bar{\boldsymbol{b}}_{n}=\mu_{n}\boldsymbol{b}_{n}$\tabularnewline
\hline 
$\boldsymbol{c}_{n}=f(\boldsymbol{x}_{n}+\boldsymbol{a}_{n})-f(\boldsymbol{x}_{n})$ & $\bar{\boldsymbol{c}}_{n}=\mu_{n}\boldsymbol{c}_{n}$\tabularnewline
\hline 
\end{tabular}
\par\end{centering}
\bigskip
\caption{\label{tab:Vector-variables-for}Vector variables for parameter-free
and parametric Steffensen's method applied to image reverse filtering.}

\end{table}
 The difference in the implementation of parametric-free and parametric
Steffensen's methods can be demonstrated by using method $\mathcal{A}1$
as an example. Referring to Table \ref{tab:A-family-of}, we can see
for both parameter-free and parametric methods the parameter $\lambda_{n}$
does not depend on $\mu_{n}$. This is evident by examining the parameter:
\begin{equation}
\lambda_{n}=\frac{||\bar{\boldsymbol{a}}_{n}||^{2}}{\bar{\boldsymbol{a}}_{n}^{T}(\bar{\boldsymbol{a}}_{n}-\bar{\boldsymbol{b}}_{n})}=\frac{||\boldsymbol{a}_{n}||^{2}}{\boldsymbol{a}_{n}^{T}(\boldsymbol{a}_{n}-\boldsymbol{b}_{n})}.\label{eq:A2lambda}
\end{equation}
where $\mu_{n}$ in numerator and denominator cancels out. Therefore,
both iteration methods can be written in the same form as the following
\begin{equation}
\boldsymbol{x}_{n+1}=\boldsymbol{x}_{n}+\mu_{n}\lambda_{n}\boldsymbol{a}_{n}\label{eq:A1}
\end{equation}
where $\mu_{n}$ is defined by one of the methods stated in section
\ref{subsec:Exponential-decay-and}. For the parameter-free Steffensen's
method, we set $\mu_{n}=1$.

In our experiments, we find that the absolute value of $\lambda_{n}$
can be very large for some filters, resulting unstable iterations.
To tackle this problem, we use a hard-limiter defined as
\begin{equation}
\hat{\lambda}_{n}=\text{sign}(\lambda_{n})\min(\tau,|\lambda_{n}|)\label{eq:hl}
\end{equation}
where $\tau$ is a user define upper limit for $|\lambda_{n}|$. A
hard-limiter is one of the essential tools in robust signal processing
\cite{RobustSigProc}. In light of this modification, we can rewrite
(\ref{eq:A1}) in the following unified way
\begin{equation}
\boldsymbol{x}_{n+1}=\boldsymbol{x}_{n}+\mu_{n}\hat{\lambda}_{n}\boldsymbol{a}_{n}\label{eq:A1-1}
\end{equation}
A parameter-free Steffensen's method is one with $\tau=\infty$ and
$\mu_{n}=1$.

To illustrate the implementation of the parametric (using Chebyshev
sequence) Steffensen's method for iteration scheme $\mathcal{A}1$
with Nesterov acceleration, necessary steps are listed in Algorithm
\ref{alg:algo}. Other iteration schemes can be similarly implemented.\begin{algorithm}[!t]
  \KwInput{$f:\:$ the black-box filter \newline $\boldsymbol{x}_0:\:$ observed image\newline $\tau=0.75$ and $P=64:\:$ two user defined parameters \newline $N:\:$ number of iterations }   
\KwOutput{$\boldsymbol{x}$}   
$n=0$\\
$t_0=1$\\
$\boldsymbol{u}_0=\boldsymbol{x}_0$\\
$\boldsymbol{x}=\boldsymbol{x}_0$\\   
\While{ $n < N$}{    	
$\boldsymbol{a}=\boldsymbol{x}_{0}-f(\boldsymbol{x})$\\
$\boldsymbol{b}=\boldsymbol{x}_{0}-f(\boldsymbol{x}+\boldsymbol{a})$\\
$\boldsymbol{c}=\boldsymbol{a}-\boldsymbol{b}$\\
$\lambda = \boldsymbol{a}^T\boldsymbol{a}/\boldsymbol{a}^T\boldsymbol{c}$\\
$\lambda = \text{sign}(\lambda)\min(\tau,|\lambda|)$\\
$\mu = 2\min(1,1/(1+\cos(1/(2\pi(n+1)/P))))$\\
$\boldsymbol{u}_1=\boldsymbol{x} + \mu\lambda\boldsymbol{a}$\\
$t_1 = (1 + \sqrt{1+4t_0^2})/2$\\
$\boldsymbol{x} = \boldsymbol{u}_1 + ((t_0-1)/t_1)(\boldsymbol{u}_1 - \boldsymbol{u}_0)$\\     
$t_0 = t_1$\\
$\boldsymbol{u}_0 = \boldsymbol{u}_1$\\
$n=n+1$
} \caption{$\mathcal{A}1$ iteration with Chebyshev sequence and  Nesterov acceleration} \label{alg:algo} \end{algorithm}

\subsubsection{Complexity}

We now comment on the computational complexity. There are two main
operations in the proposed iterations: (1) vector operations such
as inner product and scalar-vector multiplication, and (2) two calls
of the black-box filter in each iteration. We can see that the complexity
of the vector operation is $O(n)$ and the complexity of the black-box
filter depends on the nature of the filter and the particular implementation.
For example, complexity of a nonlinear filter is usually higher than
that of a linear filter, which is $O(n)$. Therefore, the complexity
of the iterations depends on the complexity of the black-box filter.

\subsubsection{Convergence}

The iteration schemes discussed in this work have a general form of
fixed-point iteration $\boldsymbol{x}_{n+1}=K(\boldsymbol{x}_{n})$.
The condition for an iteration to be convergent \cite{kelley1995iterative}
is that the function $K:\Omega\rightarrow R^{N}$ ($\Omega\subset R^{N}$)
is a contraction mapping which is define as for $\boldsymbol{x},\boldsymbol{y}\in\Omega$
\begin{equation}
||K(\boldsymbol{x})-K(\boldsymbol{y})||\le\gamma||\boldsymbol{x}-\boldsymbol{y}||\label{eq:lipsch1}
\end{equation}
where $||.||$ is a norm on $R^{N}$ and $\gamma<1$. Because the
highly nonlinear nature of the iteration schemes, it is difficult
to perform an analytical study to see if the iteration function $K(\boldsymbol{x})$
is a contraction. For example, to test if the iteration function of
the parametric scheme $\mathcal{A}1$ is a contraction, we can use
any two images $\boldsymbol{I}$ and $\boldsymbol{J}$ and write 
\begin{equation}
K(\boldsymbol{x})=\boldsymbol{x}+\hat{\lambda}(\boldsymbol{x})(\boldsymbol{x}_{0}-f(\boldsymbol{x}))\label{eq:lipsch2}
\end{equation}
and
\begin{equation}
K(\boldsymbol{y})=\boldsymbol{y}+\hat{\lambda}(\boldsymbol{y})(\boldsymbol{y}_{0}-f(\boldsymbol{y}))\label{eq:lipsch3}
\end{equation}
where $\boldsymbol{x}_{0}=f(\boldsymbol{I})$, $\boldsymbol{y}_{0}=f(\boldsymbol{J})$,
and $\hat{\lambda}(\boldsymbol{x})$ and $\hat{\lambda}(\boldsymbol{y})$
are defined by equation (\ref{eq:hl}). If $K$ is a contraction mapping,
then the following condition should be satisfied 
\begin{equation}
\frac{||K(\boldsymbol{x}_{n})-K(\boldsymbol{y}_{m})||}{||\boldsymbol{x}_{n}-\boldsymbol{y}_{m}||}<1\label{eq:lipsch4}
\end{equation}
for $m,n\ge1$. The difficulty in proving/disproving this condition
for a particular filter is due to (1) the condition depends on the
filter which can be nonlinear and usually does not have a close-form
representation, and (2) $\hat{\lambda}$ is non-linearly depended
on the filter and current iteration results. Therefore, we will only
study the convergence from numerical results which are presented in
Section \ref{sec:Results}.

\subsection{Discussion}

\subsubsection{Generalization of previous works}

Previous methods such as the T-method \cite{Tmethod} and TDA-method
\cite{TDAmethod} effectively simplify the problem by ignoring the
interactions between pixels. For example, the output of an average
filter at location $p$ depends on pixels in the neighborhood of $p$.
Using such a simplification, each pixel is treated as independent
to each other. The P-method and the S-method \cite{Pmethod} take
into consideration of the interaction by introducing a scalar parameter
which is depended on the norm of the image. However, in every iteration,
each pixel is also treated as independent. The proposed Steffensen
iteration schemes are in similar forms as those of the P-method and
the S-method. The difference is that the proposed methods are explicitly
derived from vectorization of the scalar variable form of Steffensen's
method by using notion of vector inverse defined by Brezinski and
geometric product. 

We now comment on how some of the proposed algorithms are generalization
of previous works. For example, the T-method and the TDA-method can
be written as
\begin{equation}
\text{T}:\boldsymbol{x}_{n+1}=\boldsymbol{x}_{n}+\boldsymbol{a},\label{eq:T}
\end{equation}
and
\begin{equation}
\text{TDA: }\boldsymbol{x}_{n+1}=\boldsymbol{x}_{n}+\boldsymbol{c}.\label{eq:TDA}
\end{equation}
They can be regarded as special cases of the $\mathcal{A}$-type of
iterations ($\mathcal{A}1-\mathcal{A}3$) and the $\mathcal{C}$-type
of iteration $\mathcal{C}1$, respectively. In addition, the S-method
can be written as 
\begin{equation}
\boldsymbol{x}_{n+1}=\boldsymbol{x}_{n}+\frac{||\boldsymbol{a}||_{M}}{||\boldsymbol{c}||_{M}}\boldsymbol{c}\label{eq:S}
\end{equation}
where the signal is treated as a matrix and the notation $||.||_{M}$
is used to represent the matrix 2-norm to avoid confusion with the
vector Euclidean norm. Comparing the iteration of the S-method with
that of $\mathcal{C}1$, we can see that only difference is the way
the norm is calculated. While the complexity of matrix 2-norm is $O(n^{2})$,
the complexity of the vector Euclidean norm is $O(n)$. In addition,
this work also derives two more iterations ($\mathcal{C}2$ and $\mathcal{C}3$)
that use $\boldsymbol{c}$ in the update term. 

\subsubsection{Difficulty in applying quasi-Newton methods}

In light of solving the semi-blind reverse filtering problem as solving
a system of nonlinear equations, one may attempt to use quasi-Newton's
methods such as Broyden's method \cite{kelley1995iterative} given
by the following iterations: 
\begin{equation}
\boldsymbol{x}_{n+1}=\boldsymbol{x}_{n}-\boldsymbol{J}_{n}g(\boldsymbol{x}_{n})\label{eq:broyden}
\end{equation}
The inverse of the Jacobian matrix given by
\begin{equation}
\boldsymbol{J}_{n}=\boldsymbol{J}_{n-1}+\frac{\Delta\boldsymbol{x}_{n}-\boldsymbol{J}_{n-1}\Delta g_{n}}{\Delta\boldsymbol{x}_{n}^{T}\boldsymbol{J}_{n-1}\Delta g_{n}}\Delta\boldsymbol{x}_{n}^{T}\boldsymbol{J}_{n-1}.\label{eq:jacobian}
\end{equation}
where $\Delta\boldsymbol{x}_{n}=\boldsymbol{x}_{n}-\boldsymbol{x}_{n-1}$
and $\Delta g_{n}=g(\boldsymbol{x}_{n})-g(\boldsymbol{x}_{n-1})$.
In each iteration, the computational costs are (1) two matrix-vector
multiplications, (2) one vector-vector outer products and one vector-vector
inner product, and (3) one call of the filter. The Broyden method
is not practical for this problem because of the huge size of the
matrix $\boldsymbol{J}$. For example, an image of size $(H,W)$ leads
to the matrix $\boldsymbol{J}$ of size $(HW,HW)$. Since current
digital cameras are of resolution of several million pixels, e.g.,
$H=4000$ and $W=6000$, the size of the matrix is $(2.4\times10^{7},2.4\times10^{7})$
which has $5.76\times10^{14}$ elements. If the single data type (4-byte/element)
is used, it requires $2.304\times10^{6}$ gigabyte of memory.

\section{Numerical examples\label{sec:Results}}

\subsection{Experiment setup}

There are 108 variants of the vector variable Steffensen's methods
studied in this work. For example, in the parameter-free case with
$\mu_{n}=1$, there are 36 variants: 12 methods shown in Table \ref{tab:A-family-of}
without acceleration, and the same set of 12 methods with Nesterov
or AFM acceleration. Similar variants can be constructed for the three
parametric cases denoted $\mu=\text{ed-1}$, $\mu=\text{ed-2}$, and
$\mu=\text{Cheby.}$ Each case has 36 variants.

To test the performance of these methods, we apply them to reverse
effects of 4 filters: self-guided filter ($r=35,$ $\epsilon=0.01$),
Gaussian filter ($\sigma=1$ and $\sigma=5$), and bilateral filter
($\sigma_{s}=3$, $\sigma_{r}=0.1$). The effects of these filters
are displayed in the second row of Fig. \ref{fig:Best-results images}.
We can see that images processed by the guided filter and the Gaussian
filter with $\sigma=1$ retain more information than images processed
by the Gaussian filter with $\sigma=5$ and the bilateral filter.
Since the filtered images are of different levels of peak-signal-to-noise
ratio (PSNR), the performance of reverse filtering methods is measured
by the percentage increase in PSNR
\[
p_{_{n}}=\frac{\text{PSNR}_{n}-\text{PSNR}_{0}}{\text{PSNR}_{0}}\times100\%
\]
where $n$ is the iteration index and $\text{PSNR}_{0}$ is the PSNR
of the observed image $\boldsymbol{x}_{0}$.

\subsection{Results of best performing methods}

We have conducted comprehensive experiments to evaluate all 108 variants
of the proposed vector variable Steffensen's method. Due to space
limitation, all results and analysis are presented in the Supplementary
Material. From these results, we observe that among the tested methods,
$\mathcal{{A}}{4}$, $\mathcal{{B}}{4}$, and $\mathcal{{C}}{3}$
are the most effective for reversing the effects of two filters, namely
Gaussian with $\sigma=1$ and guided filter, which are relatively
easy to reverse. For the other two filters, namely Gaussian with $\sigma=5$
and bilateral filter, which are more difficult to reverse, the top-performing
methods are $\mathcal{{A}}{1}$, $\mathcal{{B}}{1}$, and $\mathcal{{C}}{1}$.
Thus, these methods have been chosen as representative methods in
each type ($\mathcal{{A}},\mathcal{{B}}$, $\mathcal{{C}}$) for reversing
the effect of each filter. Our aim is to use these methods as examples
to highlight what can be achieved by each type of vector Steffensen's
methods and their variants.

In Fig. \ref{fig:three graphs}, we compare the performance of these
methods and their variants. A summary of the PSNR improvement properties
is listed in Table \ref{tab:convergence}. We can make the following
observations.
\begin{enumerate}
\item The first row of the four sub-figures in Fig. \ref{fig:three graphs}
displays the results obtained without further acceleration by using
either Nesterov or AFM methods. Among the tested methods, the three
parametric Steffensen's methods generally converge faster than their
parameter-free counterparts in reversing the effects of a Gaussian
filter with $\sigma=1$ and the guided filter. However, for the Gaussian
filter with $\sigma=5$ and the bilateral filter, the results are
mixed. For instance, although setting $\mu_{n}$ as the Chebyshev
sequence yields the best results for most cases, it leads to a divergent
result in reversing the effect of the bilateral filter using the method
$\mathcal{{B}}_{1}$, which converges in the parameter-free case.
Additionally, we observe that setting $\mu_{n}$ to \textquotedbl ed-1\textquotedbl{}
or \textquotedbl ed-2\textquotedbl{} can enhance the convergence
speed of the method $\mathcal{{C}}_{1}$ for both filters. Nevertheless,
this setting may result in a divergent or inferior outcome compared
to that of the corresponding parametric methods $\mathcal{{A}}_{1}$
and $\mathcal{{B}}_{1}$. Compared with the parametric counterpart,
the parameter-free methods improve the PSNR all 4 filters at the cost
of relatively slower speed.
\item The results obtained with Nesterov and AFM acceleration are shown
in the second and third rows of the four sub-figures in Fig. \ref{fig:three graphs}.
Generally, Nesterov acceleration outperforms AFM acceleration in terms
of the number of cases of PSNR improvement, as summarized in Table \ref{tab:convergence}
(b) and (c). For example, when reversing the effect of the guided
filter, Nesterov acceleration improves the convergence speed of both
parametric and parameter-free methods, albeit with a slight reduction
in the percentage increase in PSNR compared to the case without acceleration,
as demonstrated in the second row of Fig. \ref{fig:three graphs}
(c). On the other hand, the AFM acceleration does not lead to satisfactory
results, as shown in the third row of Fig. \ref{fig:three graphs} (c).
\item The parameter-free method without further acceleration is less likely
to diverge. This is evident for the two difficult filters: Gaussian
filter with $\sigma=5$ and the bilateral filter. Using Nesterov or
AFM acceleration with both parametric and parameter-free methods can
lead to divergent results in some cases, although the corresponding
parameter-free methods converge. The cause of the problem can be shown
as follows. When the iteration index $n$ is sufficiently large, the
iteration $\boldsymbol{x}_{n+1}=\varphi(\boldsymbol{x}_{n})$ using
Nesterov acceleration has the following two steps:
\begin{equation}
\boldsymbol{u}_{n+1}=\varphi(\boldsymbol{x}_{n})\label{eq:nest1}
\end{equation}
and 
\begin{equation}
\boldsymbol{x}_{n+1}=2\boldsymbol{u}_{n+1}-\boldsymbol{u}_{n}\label{eq:nest2}
\end{equation}
which can be combined into one step
\begin{equation}
\boldsymbol{u}_{n+1}=\varphi(2\boldsymbol{u}_{n}-\boldsymbol{u}_{n-1})\label{eq:nest3}
\end{equation}
Experimental results show that the convergence of the iteration $\boldsymbol{x}_{n+1}=\varphi(\boldsymbol{x}_{n})$
does not guarantee the convergence of (\ref{eq:nest3}). This issue
deserves further investigation.
\item The performance of the parametric method with either Nesterov or AFM
acceleration improves in some cases as the iteration progresses, as
indicated by the increase in percentage PSNR improvement. However,
there comes a point where the percentage PSNR improvement starts to
decrease. This is also a problem for some parametric methods. Thus,
further study is required to identify when to stop the iteration before
the method's performance deteriorates. To this end, non-referenced
image quality assessment methods \cite{Hosu2019KonIQ10kAE, Zhang2019BlindIQ}
can be used to monitor the method's performance.
\end{enumerate}

\begin{centering}
    
\renewcommand{\arraystretch}{1.5}
\begin{table}

\begin{subtable}[h]{0.7\textwidth}
\begin{centering}
\tabcolsep=0.072cm
\begin{tabular}{|c|c|c|c|c|c|c|c|c|c|c|c|c|}
\hline 
method & \multicolumn{4}{c|}{$\mathcal{{A}}_{4}$} & \multicolumn{4}{c|}{$\mathcal{{B}}_{4}$} & \multicolumn{4}{c|}{$\mathcal{{C}}_{3}$}\tabularnewline
\hline 
$\mu$ & 1 & ed-1 & ed-2 & Cheby & 1 & ed-1 & ed-2 & Cheby & 1 & ed-1 & ed-2 & Cheby\tabularnewline
\hline 
Gauss-1 & I & I & I & I & I & I & I & I & I & I & I & I\tabularnewline
\hline 
Guided & I & I & I & I & I & I & I & I & I & I & I & I\tabularnewline
\hline 
\hline 
method & \multicolumn{4}{c|}{$\mathcal{{A}}_{1}$} & \multicolumn{4}{c|}{$\mathcal{{B}}_{1}$} & \multicolumn{4}{c|}{$\mathcal{{C}}_{1}$}\tabularnewline
\hline 
$\mu$ & 1 & ed-1 & ed-2 & Cheby & 1 & ed-1 & ed-2 & Cheby & 1 & ed-1 & ed-2 & Cheby\tabularnewline
\hline 
Gauss-5 & I & I & I & I & I & I/F & I/F & I & I & I & I & I\tabularnewline
\hline 
Bilateral & I & I & I & I & I & I/F & I/F & I/F & I & I & I & I\tabularnewline
\hline 
\end{tabular}
\vspace*{0.2cm}
\caption{Results obtained without Nesterov or AFM acceleration.}
\par\end{centering}
\end{subtable}
\vspace*{0.5cm}

\begin{subtable}[h]{0.7\textwidth}
\begin{centering}
\tabcolsep=0.057cm
\begin{tabular}{|c|c|c|c|c|c|c|c|c|c|c|c|c|}
\hline 
method  & \multicolumn{4}{c|}{$\mathcal{{A}}_{4}$} & \multicolumn{4}{c|}{$\mathcal{{B}}_{4}$} & \multicolumn{4}{c|}{$\mathcal{{C}}_{3}$}\tabularnewline
\hline 
$\mu$ & 1 & ed-1 & ed-2 & Cheby & 1 & ed-1 & ed-2 & Cheby & 1 & ed-1 & ed-2 & Cheby\tabularnewline
\hline 
Gauss-1 & I & I & I & I/F & I & I & I & I/F & I & I & I & I/F\tabularnewline
\hline 
Guided & I & I & I & I & I & I & I & I & I & I & I & I\tabularnewline
\hline 
\hline 
method & \multicolumn{4}{c|}{$\mathcal{{A}}_{1}$} & \multicolumn{4}{c|}{$\mathcal{{B}}_{1}$} & \multicolumn{4}{c|}{$\mathcal{{C}}_{1}$}\tabularnewline
\hline 
$\mu$ & 1 & ed-1 & ed-2 & Cheby & 1 & ed-1 & ed-2 & Cheby & 1 & ed-1 & ed-2 & Cheby\tabularnewline
\hline 
Gauss-5 & I & F & F & F & I & F & F & F & I/F & I/F & I/F & F\tabularnewline
\hline 
Bilateral & I & I & I & I & I/F & I/F & I/F & I/F & I & I & I & I/F\tabularnewline
\hline 
\end{tabular}
\vspace*{0.2cm}
\caption{Results obtained by using Nesterov acceleration.}
\par\end{centering}
\end{subtable}
\vspace*{0.5cm}

\begin{subtable}[h]{0.7\textwidth}
\begin{centering}
\tabcolsep=0.052cm
\begin{tabular}{|c|c|c|c|c|c|c|c|c|c|c|c|c|}
\hline 
method| & \multicolumn{4}{c|}{$\mathcal{{A}}_{4}$} & \multicolumn{4}{c|}{$\mathcal{{B}}_{4}$} & \multicolumn{4}{c|}{$\mathcal{{C}}_{3}$}\tabularnewline
\hline 
$\mu$ & 1 & ed-1 & ed-2 & Cheby & 1 & ed-1 & ed-2 & Cheby & 1 & ed-1 & ed-2 & Cheby\tabularnewline
\hline 
Gauss-1 & I & I & I/F & F & I & I/F & I/F & I & I & I & I & I/F\tabularnewline
\hline 
Guided & I & I/F & I/F & I/F & I/F & I/F & I/F & I & I & I/F & I/F & I/F\tabularnewline
\hline
\hline 
method & \multicolumn{4}{c|}{$\mathcal{{A}}_{1}$} & \multicolumn{4}{c|}{$\mathcal{{B}}_{1}$} & \multicolumn{4}{c|}{$\mathcal{{C}}_{1}$}\tabularnewline
\hline 
$\mu$ & 1 & ed-1 & ed-2 & Cheby & 1 & ed-1 & ed-2 & Cheby & 1 & ed-1 & ed-2 & Cheby\tabularnewline
\hline 
Gauss-5 & F & F & F & F & I & F & F & F & I/F & I/F & I/F & F\tabularnewline
\hline 
Bilateral & I/F & I/F & I/F & I/F & I/F & I/F & I/F & I/F & I & I/F & I/F & I\tabularnewline
\hline 
\end{tabular}
\vspace*{0.2cm}
\caption{Result obtained from using AFM acceleration}
\par\end{centering}
\end{subtable}
\bigskip
\caption{\label{tab:convergence}Summary of PSNR improvement for all cases presented in Fig. \ref{fig:three graphs}.  "I" means improvement, "F" means no improvement or divergent, and "I/F" means unstable performance.}
\end{table}
\end{centering}
\begin{figure*}[h!]
     \centering
     \begin{subfigure}[b]{0.495\textwidth}
         \centering
         \includegraphics[viewport=50bp 0bp 970bp 1010bp,clip,width=\textwidth]{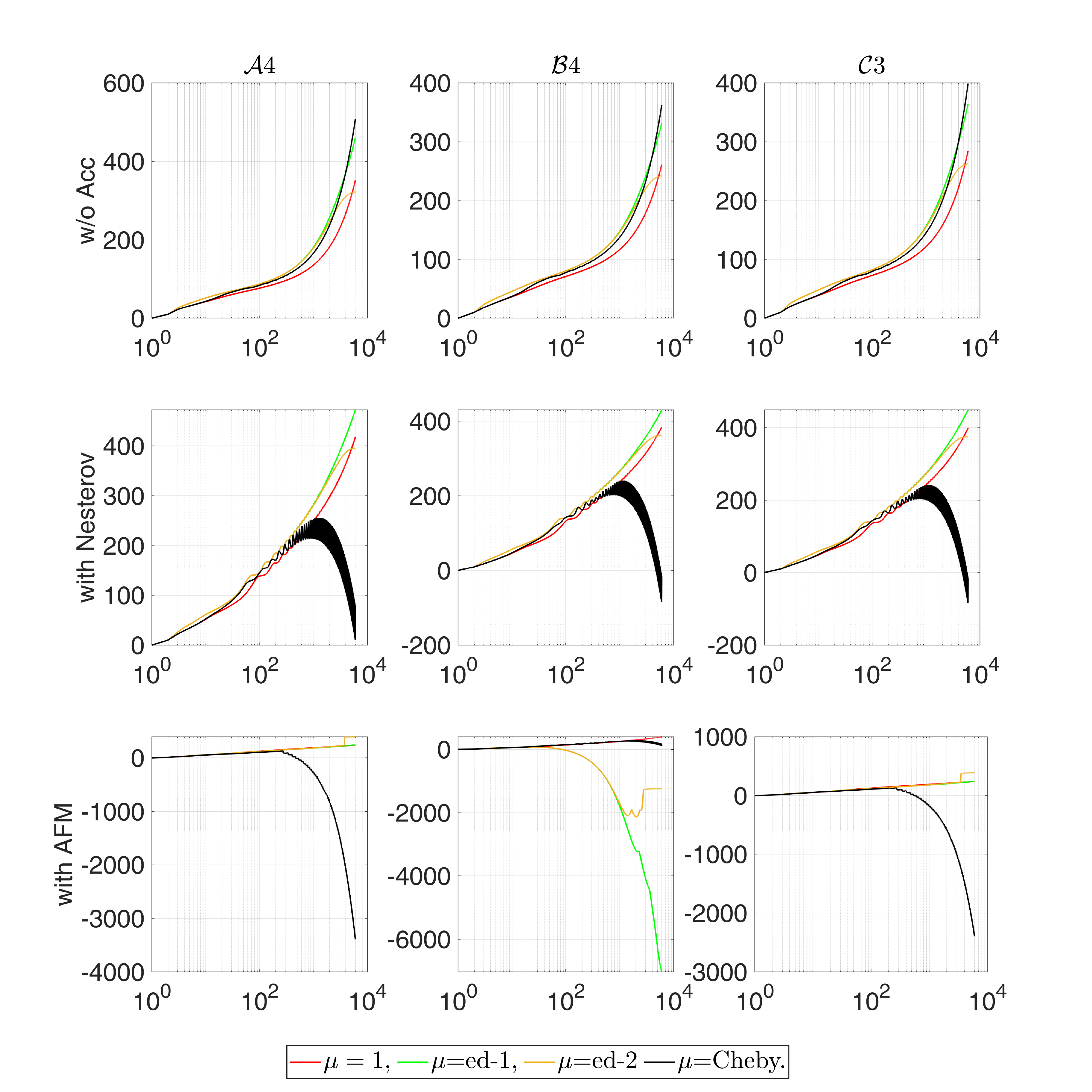}
         \caption{Gaussian $\sigma=1$.}
     \end{subfigure}
     \begin{subfigure}[b]{0.495\textwidth}
         \centering
         \includegraphics[viewport=50bp 0bp 970bp 1010bp,clip, width=\textwidth]{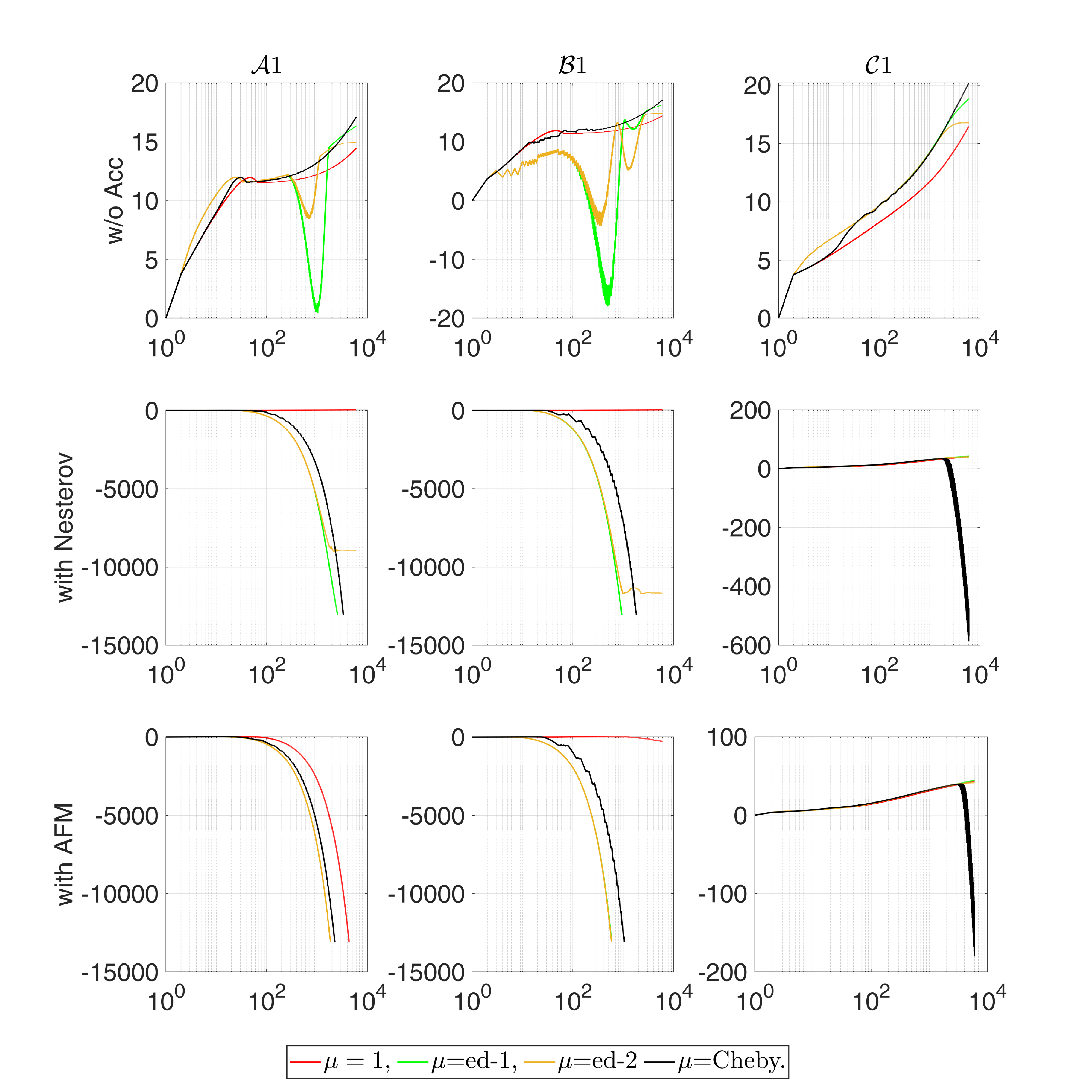}
         \caption{Gaussian $\sigma=5$.}
     \end{subfigure}
\bigskip
 \begin{subfigure}[b]{0.495\textwidth}
         \centering
         \includegraphics[viewport=50bp 0bp 970bp 1010bp,clip,width=\textwidth]{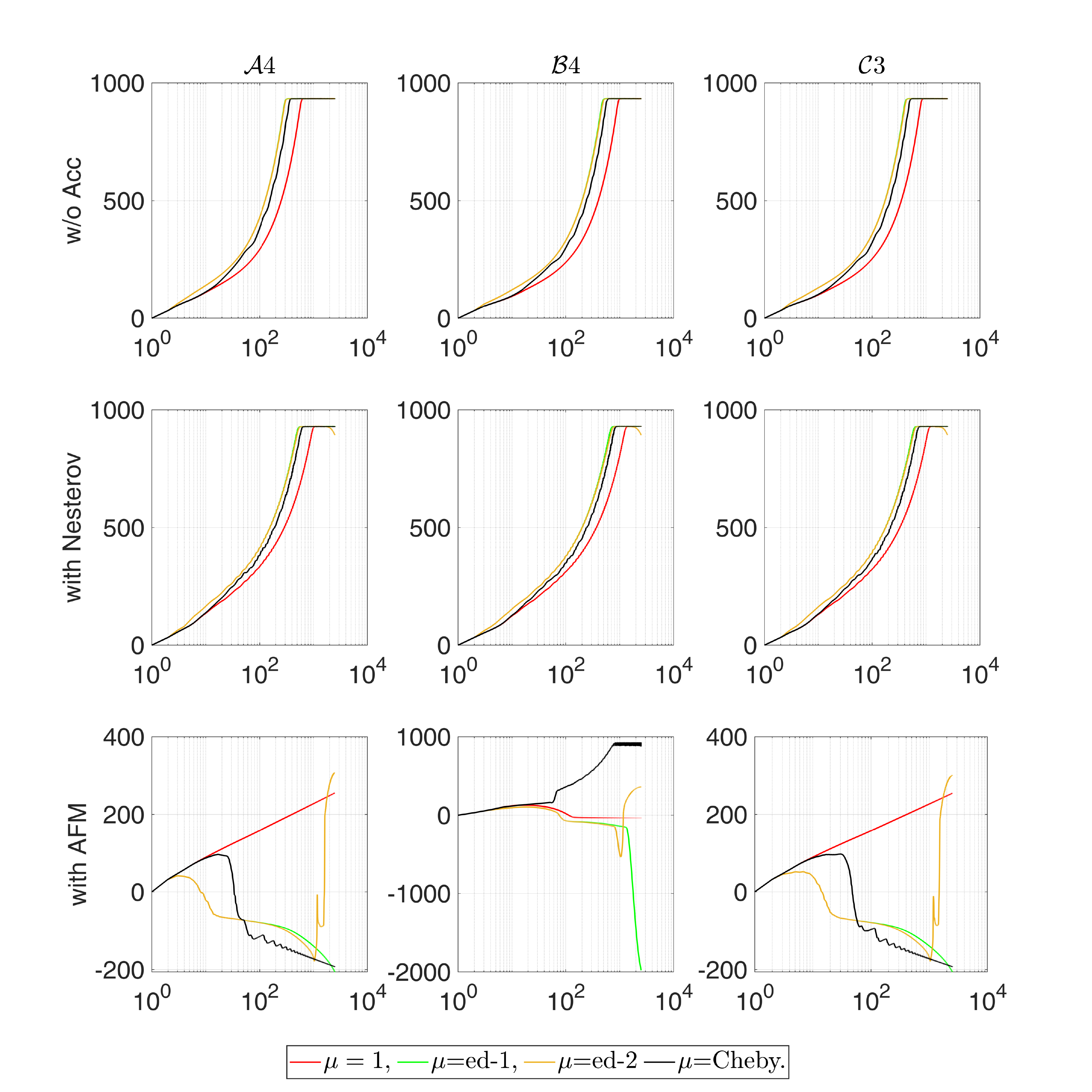}
         \caption{Guided filter.}
     \end{subfigure}
     \begin{subfigure}[b]{0.495\textwidth}
         \centering
         \includegraphics[viewport=50bp 0bp 970bp 1010bp,clip,width=\textwidth]{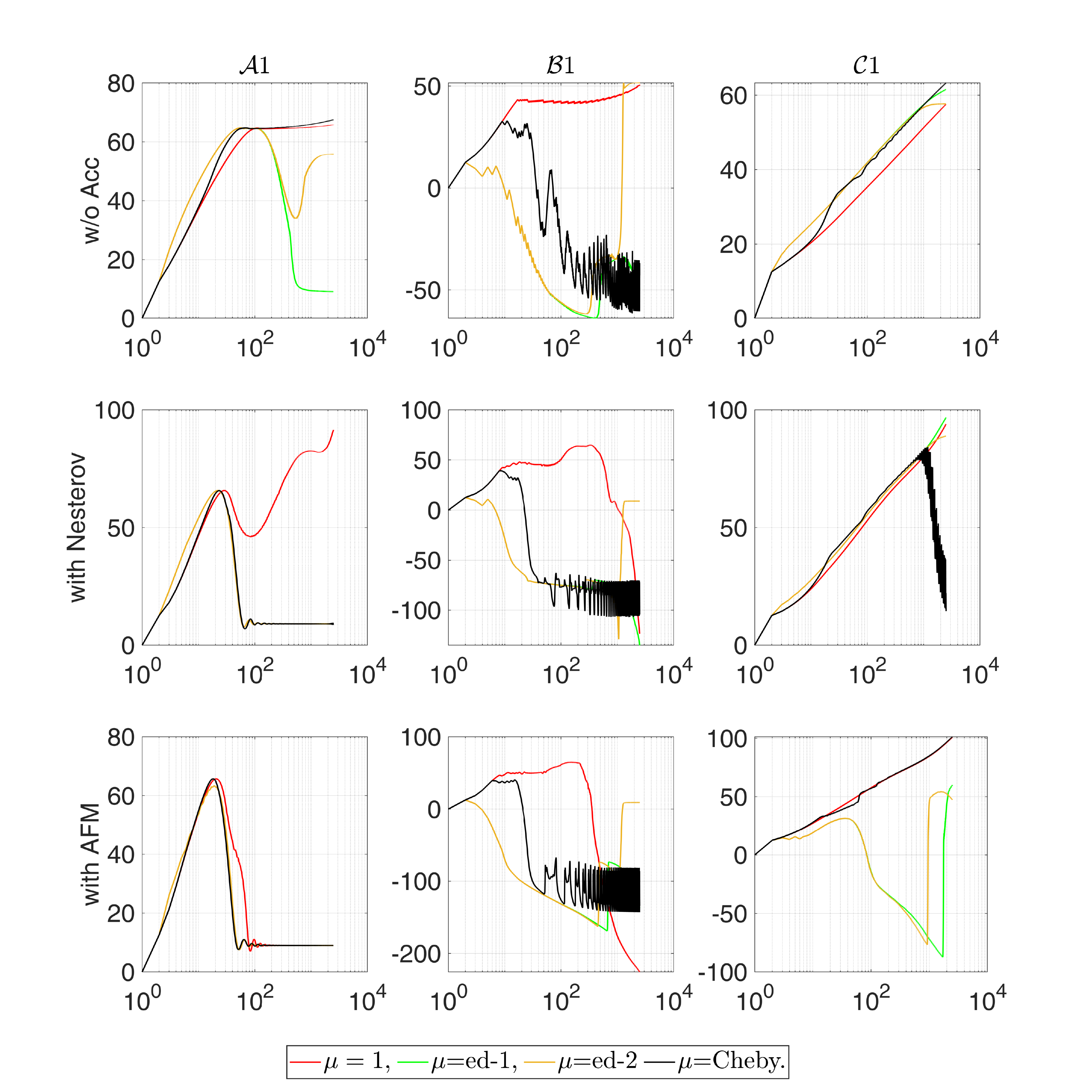}
         \caption{ Bilateral filter.}
     \end{subfigure}
        \caption{Comparison of the best performing methods in each category ($\mathcal{A}$, $\mathcal{B}$, $\mathcal{C}$) in inverting effects of 4 filters  (a) Gaussian $\sigma=1$, (b) Gaussian $\sigma=5$, (c) Guided filter, and (d) Bilateral filter.  Horizontal and vertical axis are the iteration index and percentage improvement of PSNR, respectively.}
        \label{fig:three graphs}
\end{figure*}

\subsection{Comparison with best performing existing methods\label{subsec:Comparison-with-best}}

\begin{figure}
\begin{centering}
\includegraphics[width=0.5\columnwidth]{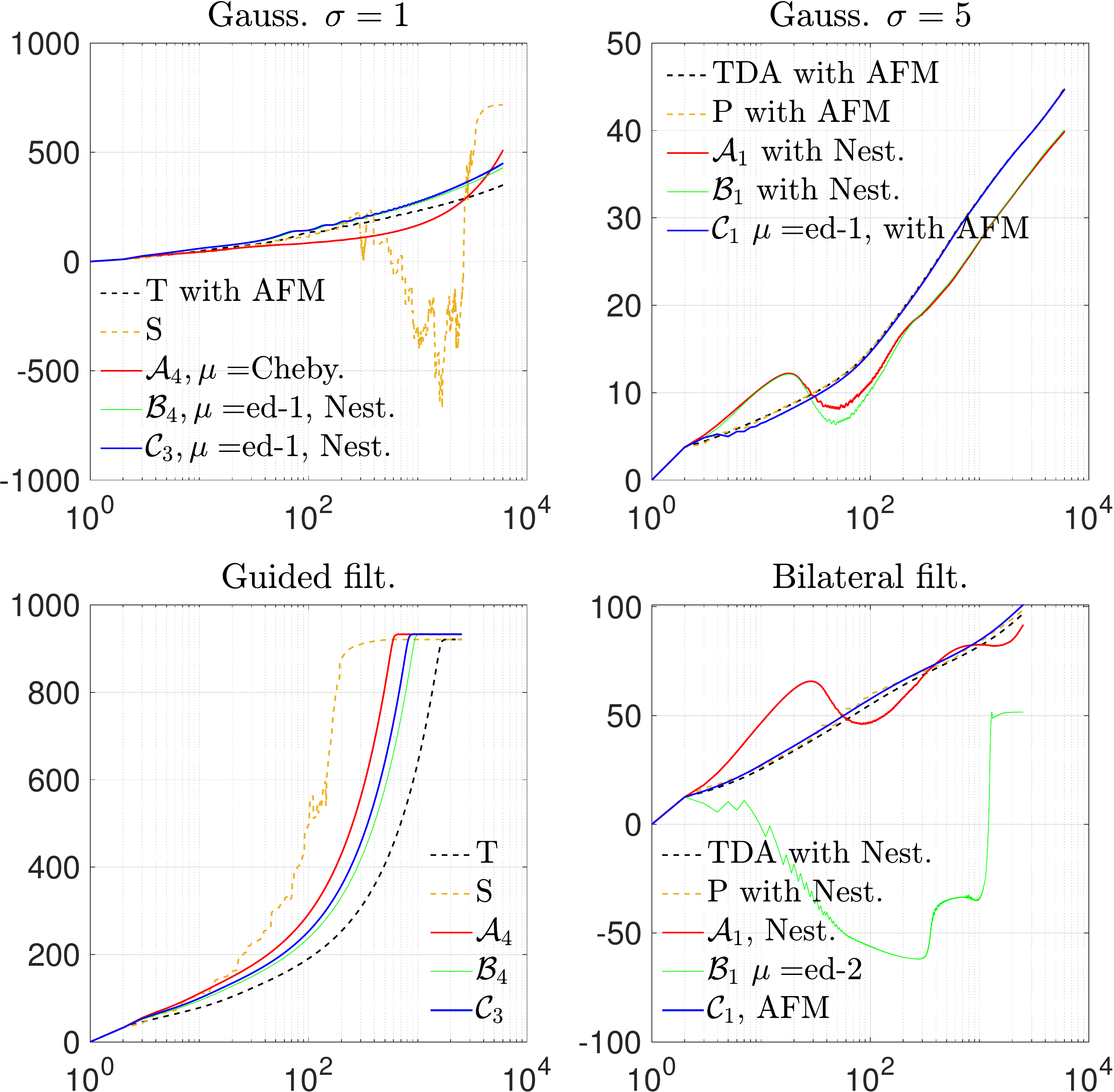}
\par\end{centering}
\caption{\label{fig:Comparison-of-the-4}Comparison of the best performing
vector variable Steffensen's method with two best performing existing
methods in inverting the effects of four filters. Horizontal and vertical
axis are the iteration index and percentage improvement of PSNR, respectively..}
\end{figure}

\begin{figure*}
\begin{centering}
\includegraphics[width=\columnwidth]{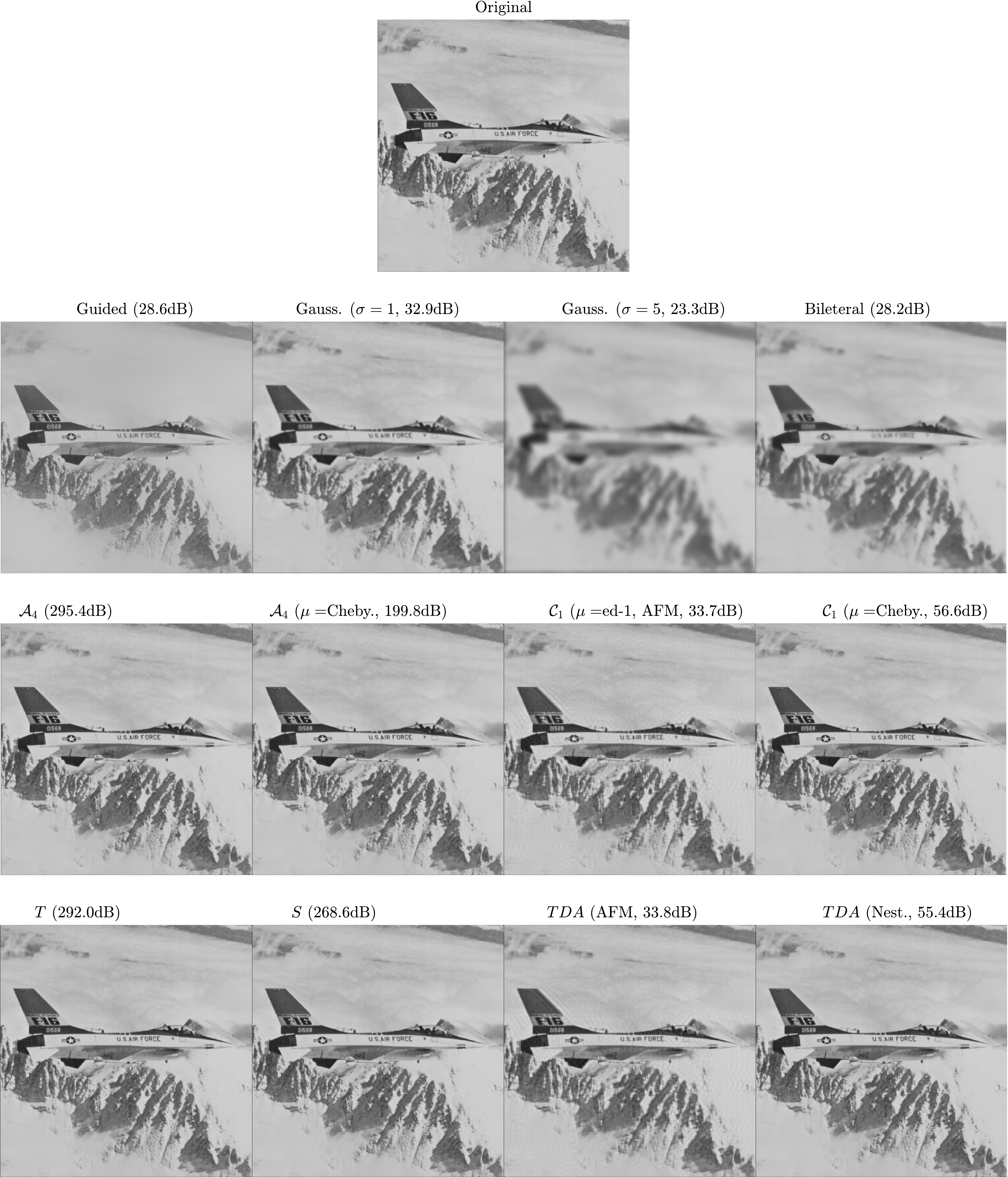}
\par\end{centering}
\caption{\label{fig:Best-results images}Comparison of best results in terms
of achieving the highest PSNR when the method reaches the preset maximum
number of iterations. The methods without specification of $\mu$
are with the setting $\mu=1$. The number in bracket, e.g., (28.5dB),
indicates the PSNR of the image below. Nesterov acceleration is abbreviated
as Nest.. Second row: images to be processed. Third row: best results
of Steffensen's method. Fourth row: best results of existing methods.}
\end{figure*}

To enable easy comparison with established techniques such as T-method
\cite{Tmethod}, TDA-method \cite{TDAmethod}, P-method and S-method
\cite{Pmethod}, we select the top two performing methods. These methods
are highlighted in Table \ref{tab:Best-performing-methods.}, chosen
based on their performance as detailed in the Supplementary Material.
The criterion for the best performance is the highest percentage improvement
in PSNR for all methods in reversing the effect of a filter at the
end of a preset number of iterations. 
\begin{table*}[t]
\begin{centering}
\begin{tabular}{|c|c|c|c||c|c|}
\hline 
 & \multicolumn{3}{c||}{Vector variable Steffensen's method} & \multicolumn{2}{c|}{T, TDA, P, S}\tabularnewline
\hline 
\hline 
Gauss, $\sigma=1$ & $\mathcal{{A}}_{4},\mu=$Cheby., (508) & $\mathcal{{B}}_{4},\mu=$ed-1, Nest., (430) & $\mathcal{{C}}_{3},\mu=$ed-1, Nest., (449) & T, AFM (350) & S (717)\tabularnewline
\hline 
Gauss. $\sigma=5$ & $\mathcal{{A}}_{1},$ Nest., (40) & $\mathcal{{B}}_{1},$Nest., (40) & $\mathcal{{C}}_{1},\mu=$ed-1, AFM., (45) & TDA, AFM, (45) & P, AFM, (45)\tabularnewline
\hline 
Guided & $\mathcal{{A}}_{4}$ (933) & $\mathcal{{B}}_{4}$ (933) & $\mathcal{{C}}_{3}$ (933) & T (921) & S (920)\tabularnewline
\hline 
Bilateral & $\mathcal{{A}}_{1},$Nest., (91) & $\mathcal{{B}}_{1},\mu=$ed-2, (52) & $\mathcal{{C}}_{1},$AFM, (101) & TDA, Nest., (97) & P, Nest., (98)\tabularnewline
\hline 
\end{tabular}
\par\end{centering}
\caption{\label{tab:Best-performing-methods.}Best performing methods chosen
based solely on the highest percentage of PSNR improvement when the
method stops at the maximum number of preset iterations. Methods without
specification of $\mu_{n}$ are with the setting $\mu_{n}=1$. The
number in bracket, e.g., (508), indicates the 508\% PSNR improvement.
Nesterov acceleration is abbreviated as ``Nest.''.}
\end{table*}
Comparison results are shown in Figures \ref{fig:Comparison-of-the-4}
and \ref{fig:Best-results images}. We have the following observations.
\begin{enumerate}
\item For the Gaussian filter with $\sigma=1$, the performance of the S-method
is unstable, but it has managed to converge to the highest improvement
in PSNR. The three vector Steffensen's methods have similar speed
of improvement and is faster than that of the T-method. Except the
S-method, the performance of all methods is on an increasing trajectory,
which suggests potentially higher degree of improvement if more iterations
are performed.
\item For the Gaussian filter with $\sigma=5$, at the end of the preset
number of iterations, all methods have achieved PSNR improvement of
about 40\%, which is relatively small compared to the previous case.
The TDA-method and the $\mathcal{C}_{1}$ method have a faster speed
of improvement than that of other methods. The improvement of PSNR
for all methods is trending upwards, indicating a possibility of further
improvement if the iteration continues.
\item For the guided filter case, all methods converge to a similar level
of PSNR improvement. The S-method has the fastest speed of convergence.
The three Steffensen's methods have similar speeds which are faster
than that of the T-method. However, we note that in terms of computational
complexity, the T-method is lowest, while the S-method is the highest.
\item For the bilateral filter case, the $\mathcal{{A}}_{1}$ method has
the fastest speed of improvement. However, the improvement oscillates
during the iteration. The performances of the other three method:
TDA-method with Nesterov, P-method with Nesterov, and $\mathcal{{C}}_{1}$
method with AFM are similar. The speed of improvement of these method
is slower than that of the $\mathcal{{A}}_{1}$ method at the start
of the iteration. The $\mathcal{{B}}_{1}$ method is an exception,
its performance deteriorates and then improves to achieves an PSNR
improvement of about 50\%.
\item The computational complexity of iterative image reverse filtering
methods depends on two main factors: the number of calls to the black
box filter (denoted N) and the complexity of the iteration process
(denoted C). Table \ref{tab:Comparison-of-complexity} presents a
comparison of the methods examined in this section. The table also
includes the running time of each method to complete a preset number
of iterations, which is 2500 for the guided filter and 6000 for the
Gaussian filter with $\sigma=1$. The running times for the other
two filters are not shown because some methods diverge. The running
times were obtained by averaging the results of 10 runs of each method
programmed in MATLAB 2022b on a computer with an Intel i9-9900KF CPU
and 64 GB of memory. The table reveals that the running times of the
TDA method and the vector Steffensen's method are about the same and
roughly double that of the T-method. This is because the former two
methods require 2 calls to the filter in 1 iteration, while the T-method
requires only 1 call. However, the P-method and the S-method take
significantly longer time to run due to the $O(n^2)$ complexity involved
in calculating the matrix 2-norm.
\item We would like to highlight that among 12 versions of the vector variable
Steffensen's methods, four methods stand out for their superior performance:
$\mathcal{{A}}_{1}$, $\mathcal{{A}}_{4}$, $\mathcal{{B}}_{1}$,
and $\mathcal{{C}}_{1}$. Importantly, these methods are novel and
have not been published previously. 
\begin{table}
\begin{centering}
\begin{tabular}{|c|c|c|c|c|c|}
\hline 
 & T \cite{Tmethod} & TDA \cite{TDAmethod} & P \cite{Pmethod} & S \cite{Pmethod} & Steffensen\tabularnewline
\hline 
\hline 
N & 1 & 2 & 2 & 2 & 2\tabularnewline
\hline 
C & $O(n)$ & $O(n)$ & $O(n^{2})$ & $O(n^{2})$ & $O(n)$\tabularnewline
\hline 
Guided & 46.1 & 91.3 & 204.2 & 160.0 & 92.4\tabularnewline
\hline 
Gauss-1 & 5.9 & 10.2 & 163.5 & 169.0 & 11.8\tabularnewline
\hline 
\end{tabular}
\par\end{centering}
\bigskip
\caption{\label{tab:Comparison-of-complexity}Rows 2 (number of filter calls)
and 3 (complexity of vector/matrix operations) show the comparison
of complexity of iterative methods studied in section \ref{subsec:Comparison-with-best},
while rows 4 and 5 show the running time (seconds) to complete 2500
and 6000 iterations for each method to reverse the effects of a guided
filter and a Gaussian filter with $\sigma=1$, respectively.}
\end{table}
\end{enumerate}

\subsection{Summary and discussion}

Comparing the proposed parameter-free and parametric Steffensen's
methods, the performance of the former is relatively more robust than
that of the latter in that they are less likely to lead to divergent
results. However, in some cases, the parametric methods actually achieve
faster speed of PSNR improvement and thus a higher degree of improvement
at the end of the preset number of iterations. When further acceleration
methods such as the Nesterov and AFM are used, the results are mixed.
Although in some cases both acceleration methods lead to faster speed
of PSNR improvement, the Nesterov acceleration is preferred because
with it the iteration is less likely to diverge. In addition, these
two acceleration methods help to improve the performance for existing
methods (T/TDA/P/S) in the two difficult cases: a Gaussian filter
with $\sigma=5$ and a bilateral filter.

These mixed results highlight (1) the importance of a comprehensive
study of the performance of variants of the proposed Steffensen's
method conducted in this work, and (2) the importance of this work
which provides a new set of tools to tackle the reverse filtering
problem. Further study should be conducted to explain why certain
variants diverge in reversing the effect of a Gaussian filter with
$\sigma=5$. An analysis of the convergence of T/TDA method using
the Fourier transform was presented in \cite{TDAmethod}.

\section{Conclusions}

This work is motivated by current research in solving the semi-blind
image reverse filtering problem which can be formulated as a solving
a system of nonlinear equations. We focus on extending Steffensen's
method to provide new tools for this problem. The first extension
is the development of the parametric Steffensen's method which can
be regarded as an acceleration of the Mann iteration. The original
Steffensen's method is an acceleration of the Picard iteration. The
second extension is the development of a family of 12 Steffensen's
methods for vector variables based on the notion of Brezinski inverse
and the vector inverse defined by the geometric product. We have presented
variants of these methods based on three ways of adaptively setting
the parameter and using accelerated first-order method (which includes
Nesterov method as a special case) to explore the possibility of further
acceleration. To demonstrate the application to the semi-blind image
reverse filtering problem, we present implementation details, analysis
of computation complexity and convergence. We also show that some
of the proposed methods generalize existing iterative algorithms.
In the accompanying Supplementary Material, we have presented a comprehensive
study of the performance of each of 108 variants of the vector Steffensen's
methods in terms of percentage improvement of PSNR at each iteration
in reversing the effects of four commonly used filters in image processing.
Based on these results, we have presented a detailed study of the
best performing methods and compared them with the best performing
existing methods. Overall, this paper provides a family of new Steffensen's
methods which are new tools for semi-blind image reverse filtering.

\section*{Supplementary information}

We present a complete set of experimental results in the Supplementary
Material, which outlines the experimental setup, presents a comprehensive
collection of figures and tables, and provides detailed analysis of
all the outcomes.

\section*{Appendix}

Details of the vectorization of Steffensen's method by using the notion
of geometric product are shown below. 

\textbf{Case A1.1-G}

\[
\boldsymbol{\omega}_{n}=\boldsymbol{a}\boldsymbol{a}\boldsymbol{c}^{-1}=\frac{||\boldsymbol{a}||^{2}}{||\boldsymbol{c}||^{2}}\boldsymbol{c}
\]
\[
\boldsymbol{x}_{n+1}=\boldsymbol{x}_{n}+\frac{||\boldsymbol{a}||^{2}}{||\boldsymbol{c}||^{2}}\boldsymbol{c}
\]

\textbf{Case A1.2-G} 
\[
\boldsymbol{\omega}_{n}=\boldsymbol{a}\boldsymbol{c}^{-1}\boldsymbol{a}=\frac{\boldsymbol{aca}}{||\boldsymbol{c}||^{2}}=\boldsymbol{a}+\frac{||\boldsymbol{a}||^{2}\boldsymbol{b}-||\boldsymbol{b}||^{2}\boldsymbol{a}}{||\boldsymbol{c}||^{2}}
\]

\begin{align*}
\boldsymbol{x}_{n+1} & =\boldsymbol{x}_{n}+\boldsymbol{a}+\frac{||\boldsymbol{a}||^{2}\boldsymbol{b}-||\boldsymbol{b}||^{2}\boldsymbol{a}}{||\boldsymbol{c}||^{2}}\\
 & =\phi(\boldsymbol{x}_{n})+\frac{||\boldsymbol{a}||^{2}\boldsymbol{b}-||\boldsymbol{b}||^{2}\boldsymbol{a}}{||\boldsymbol{c}||^{2}}
\end{align*}

\textbf{Case A2.1-G }(same as \textbf{A1.1-G})\textbf{
\[
\boldsymbol{\omega}_{n}=\boldsymbol{abc}^{-1}=\frac{\boldsymbol{abc}}{||\boldsymbol{c}||^{2}}=\frac{2(\boldsymbol{a^{T}b})\boldsymbol{a}-||\boldsymbol{a}||^{2}\boldsymbol{b}-||\boldsymbol{b}||^{2}\boldsymbol{a}}{||\boldsymbol{c}||^{2}}
\]
\begin{align*}
\boldsymbol{x}_{n+1} & =\phi(\boldsymbol{x}_{n})+\frac{2(\boldsymbol{a^{T}b})\boldsymbol{a}-||\boldsymbol{a}||^{2}\boldsymbol{b}-||\boldsymbol{b}||^{2}\boldsymbol{a}}{||\boldsymbol{c}||^{2}}\\
 & =\boldsymbol{x}_{n}+\boldsymbol{a}+\frac{2(\boldsymbol{a^{T}b})\boldsymbol{a}-||\boldsymbol{a}||^{2}\boldsymbol{b}-||\boldsymbol{b}||^{2}\boldsymbol{a}}{||\boldsymbol{c}||^{2}}\\
 & =\boldsymbol{x}_{n}+\frac{||\boldsymbol{a}||^{2}}{||\boldsymbol{c}||^{2}}\boldsymbol{c}
\end{align*}
}

\textbf{Case A2.2-G }(same as \textbf{A1.2-G})\textbf{
\[
\boldsymbol{\omega}_{n}=\boldsymbol{a}\boldsymbol{c}^{-1}\boldsymbol{b}=\frac{\boldsymbol{acb}}{||\boldsymbol{c}||^{2}}=\frac{||\boldsymbol{a}||^{2}\boldsymbol{b}-||\boldsymbol{b}||^{2}\boldsymbol{a}}{||\boldsymbol{c}||^{2}}
\]
}

\[
\boldsymbol{x}_{n+1}=\phi(\boldsymbol{x}_{n})+\frac{||\boldsymbol{a}||^{2}\boldsymbol{b}-||\boldsymbol{b}||^{2}\boldsymbol{a}}{||\boldsymbol{c}||^{2}}
\]

\textbf{Case A2.3-G }(same as \textbf{A1.2-G})\textbf{
\[
\boldsymbol{\omega}_{n}=\boldsymbol{b}\boldsymbol{c}^{-1}\boldsymbol{a}=\frac{\boldsymbol{bca}}{||\boldsymbol{c}||^{2}}=\frac{||\boldsymbol{a}||^{2}\boldsymbol{b}-||\boldsymbol{b}||^{2}\boldsymbol{a}}{||\boldsymbol{c}||^{2}}
\]
}

\textbf{Case A2.4-G
\[
\boldsymbol{\omega}_{n}=\boldsymbol{ba}\boldsymbol{c}^{-1}=\frac{\boldsymbol{bac}}{||\boldsymbol{c}||^{2}}=\frac{||\boldsymbol{a}||^{2}\boldsymbol{b}-2(\boldsymbol{a^{T}b})\boldsymbol{b}+||\boldsymbol{b}||^{2}\boldsymbol{a}}{||\boldsymbol{c}||^{2}}
\]
}
\begin{align*}
\boldsymbol{x}_{n+1} & =\phi(\boldsymbol{x}_{n})+\frac{||\boldsymbol{a}||^{2}\boldsymbol{b}-2(\boldsymbol{a^{T}b})\boldsymbol{b}+||\boldsymbol{b}||^{2}\boldsymbol{a}}{||\boldsymbol{c}||^{2}}\\
 & =\phi(\phi(\boldsymbol{x}_{n}))-\boldsymbol{b}+\frac{||\boldsymbol{a}||^{2}\boldsymbol{b}-2(\boldsymbol{a^{T}b})\boldsymbol{b}+||\boldsymbol{b}||^{2}\boldsymbol{a}}{||\boldsymbol{c}||^{2}}\\
 & =\phi(\phi(\boldsymbol{x}_{n}))+\frac{||\boldsymbol{b}||^{2}}{||\boldsymbol{c}||^{2}}\boldsymbol{c}
\end{align*}

\textbf{Case A2.5-G }(same as case \textbf{A2.4-G})
\[
\boldsymbol{\omega}_{n}=\boldsymbol{c}^{-1}\boldsymbol{a}\boldsymbol{b}=\frac{\boldsymbol{c}\boldsymbol{a}\boldsymbol{b}}{||\boldsymbol{c}||^{2}}=\frac{||\boldsymbol{a}||^{2}\boldsymbol{b}-2(\boldsymbol{a^{T}b})\boldsymbol{b}+||\boldsymbol{b}||^{2}\boldsymbol{a}}{||\boldsymbol{c}||^{2}}
\]

\textbf{Case A2.6-G }(same as case \textbf{A1.1-G})
\[
\boldsymbol{\omega}_{n}=\boldsymbol{c}^{-1}\boldsymbol{b}\boldsymbol{a}=\frac{\boldsymbol{c}\boldsymbol{b}\boldsymbol{a}}{||\boldsymbol{c}||^{2}}=\frac{2(\boldsymbol{a}^{T}\boldsymbol{b})\boldsymbol{a}-||\boldsymbol{b}||^{2}\boldsymbol{a}-||\boldsymbol{a}||^{2}\boldsymbol{b}}{||\boldsymbol{c}||^{2}}
\]

\begin{align*}
\boldsymbol{x}_{n+1} & =\phi(\boldsymbol{x}_{n})+\frac{2(\boldsymbol{a}^{T}\boldsymbol{b})\boldsymbol{a}-||\boldsymbol{b}||^{2}\boldsymbol{a}-||\boldsymbol{a}||^{2}\boldsymbol{b}}{||\boldsymbol{c}||^{2}}\\
 & =\boldsymbol{x}_{n}+\boldsymbol{a}\frac{2(\boldsymbol{a}^{T}\boldsymbol{b})\boldsymbol{a}-||\boldsymbol{b}||^{2}\boldsymbol{a}-||\boldsymbol{a}||^{2}\boldsymbol{b}}{||\boldsymbol{c}||^{2}}\\
 & =\boldsymbol{x}_{n}+\frac{||\boldsymbol{a}||^{2}}{||\boldsymbol{c}||^{2}}\boldsymbol{c}
\end{align*}

\textbf{Case A3.1-G }(same as \textbf{A2.4-G})

\[
\boldsymbol{\omega}_{n}=\boldsymbol{bb}\boldsymbol{c}^{-1}=\frac{||\boldsymbol{b}||^{2}}{||\boldsymbol{c}||^{2}}\boldsymbol{c}
\]
\[
\boldsymbol{x}_{n+1}=\phi(\phi(\boldsymbol{x}_{n}))+\frac{||\boldsymbol{b}||^{2}}{||\boldsymbol{c}||^{2}}\boldsymbol{c}
\]

\textbf{Case} \textbf{A3.2-G }(same as \textbf{A1.2-G})
\[
\boldsymbol{\omega}_{n}=\boldsymbol{b}\boldsymbol{c}^{-1}\boldsymbol{b}=\frac{\boldsymbol{b}(\boldsymbol{a}-\boldsymbol{b})\boldsymbol{b}}{||\boldsymbol{c}||^{2}}=-\boldsymbol{b}+\frac{||\boldsymbol{a}||^{2}\boldsymbol{b}-||\boldsymbol{b}||^{2}\boldsymbol{a}}{||\boldsymbol{c}||^{2}}
\]

\begin{align*}
\boldsymbol{x}_{n+1} & =\phi(\phi(\boldsymbol{x}_{n}))-\boldsymbol{b}+\frac{||\boldsymbol{a}||^{2}\boldsymbol{b}-||\boldsymbol{b}||^{2}\boldsymbol{a}}{||\boldsymbol{c}||^{2}}\\
 & =\phi(\boldsymbol{x}_{n})+\frac{||\boldsymbol{a}||^{2}\boldsymbol{b}-||\boldsymbol{b}||^{2}\boldsymbol{a}}{||\boldsymbol{c}||^{2}}
\end{align*}

\bibliographystyle{ieeetr}  
\bibliography{references}  

\newpage
\begin{center}
    {\LARGE Supplementary material}
\end{center}

\section*{Introduction}

The purpose of this document is to present all experimental findings
that offer additional insights into the performance of the methods
examined in the main paper for the task of reversing the effects of
four commonly employed filters in image processing. For ease of comprehension,
some information is duplicated in both this document and the main
paper.

In this document, we present the test methods in section \ref{subsec:Test-methods},
and a comprehensive list of figures and tables is available in section
\ref{subsec:Summary-of-figures}. Section \ref{sec:Results} includes
a summary and analysis of all experimental outcomes, based on which
we identify the best-performing methods for inverting the effects
of the four image filters. Such methods are listed in section \ref{subsec:Convergence-and-comparison}.
Finally, all figures and tables are presented at the end of this document.

\section*{Methodology and summary of figures and tables}

\subsection*{Test methods\label{subsec:Test-methods}}

A comprehensive test has been conducted in the application of semi-blind
reverse image filtering. Three frequently used filters are employed
in this study: the Gaussian filter with $\sigma=1$ and $\sigma=5$,
 the guided filter with $r=35$ and $\epsilon=0.01$, and the bilateral
filter with $\sigma_{s}=3$ and $\sigma_{r}=0.1$. While the Gaussian
filter with $\sigma=1$ and $\sigma=5$ represent a lightly and heavily
smoothing operation, the guided filter and the bilateral filter are
two different types of edge-aware filters which smooth small-scale
details and protect edges of large scale objects from smoothing.

The aim of this study is to evaluate the performance of vector Steffensen's
methods and corresponding acceleration versions by using the Nesterov
method and adaptive first order method (AFM). The advantage of these
two methods is that they are parameter free. Vector Steffensen's methods
studied in this work are listed as follows. Each category has the
12 variants.
\begin{enumerate}
\item Parameter free methods -- results are presented in section \ref{subsec:Parameter-free-methods-and}.
\begin{enumerate}
\item $\mu=1$.
\item $\mu=1$ and Nesterov.
\item $\mu=1$ and AFM.
\end{enumerate}
\item Parametric methods -- results are presented in section \ref{subsec:Parameter-free-methods-and}
\begin{enumerate}
\item $\mu_{n}=1+\exp(-((2n)/N)^{2})$, where $n$ and $N$ are the iteration
index and total number of iterations, respectively. This method is
called exponentially decay and is abbreviated as ``ed-1''.
\item $\mu_{n}=2\exp(-((2n)/N)^{2})$, where $n$ and $N$ are the iteration
index and total number of iterations, respectively. This method is
called exponentially decay and is abbreviated as ``ed-2''.
\item $\mu_{n}=2\min(1,\frac{1}{1+\cos(2(n+1)/P)})$, where $n$ and $P=64$
are the iteration index and the period of the Chebyshev sequence,
respectively. This method is called Chebyshev sequence and is abbreviated
as ``Cheby.''
\end{enumerate}
\item Parametric methods with Nesterov and AFM acceleration -- results
are presented in section \ref{subsec:Parametric-methods-with}.
\begin{enumerate}
\item ``ed-1'' with Nesterov.
\item ``ed-1'' with AFM.
\item ``ed-2'' with Nesterov.
\item ``ed-2'' with AFM.
\item ``Cheby.'' with Nesterov.
\item ``Cheby.'' with AFM.
\end{enumerate}
\end{enumerate}
To evaluate the performance of the above methods, we also studies
existing method such as T-method, TDA-method, P-{}-method, and S-method
and their corresponding acceleration by using the Nesterov method
and the AFM method. Results are presented in section \ref{subsec:Performance-of-existing}.
A numerical comparison of the performance in improving PSNR is presented
in section \ref{subsec:Convergence-and-comparison}.

All results are obtained by processing the gray scale ``F-16'' image
of size $512\times512$. The performance is measured by the percentage
improvement in PSNR in each iteration:
\[
p_{n}=\frac{PSNR_{n}-PSNR_{n}}{PSNR_{0}}\times100\%
\]
where $PSNR_{n}$ is the peak-signal-to-noise ratio of the filtered
image and $PSNR_{n}$ is the PSNR of the inverse filtering result
at the $n$th iteration.

\subsection*{Summary of figures and tables\label{subsec:Summary-of-figures}}

Figures and Tables are summarized below in the order of their appearance.
\begin{itemize}
\item Figures \ref{fig:Results-of-Gaussian1} to \ref{fig:Results-of-bilateral}
present results of PSNR improvement (vertical axis) versus iteration
(horizontal axis) for the parametric-free methods without and with
Nesterov or AFM acceleration. Results of parametric methods without
accelerations are also presented in these figures.
\item Figures \ref{fig:Results-of-Gaussian1-1}-\ref{fig:Results-of-bilateral-1}
present results of parametric methods with accelerations.
\item Tables \ref{tab:Gauss1}-\ref{tab:MyTableLabel-4} present results
of percentage PSNR improvement at the end of the iteration for parameter-free
and parametric method, with and without acceleration. The entry with
``-'' means the method is divergent.
\item Figure \ref{fig:three graphs} presents results of the 4 existing
methods: T, TDA, P, and S.
\item Table \ref{tab:subtables} presents results of percentage PSNR improvement
at the end of the iteration for the 4 existing methods: T, TDA, P,
and S. The entry with ``-'' means the method is divergent.
\end{itemize}

\section*{Discussion of results\label{sec:Results}}

\subsection*{Evaluation of the 12 vector Steffensen's methods and their variants}

\subsubsection*{Parameter-free methods and parametric methods.\label{subsec:Parameter-free-methods-and}}

\begin{figure}[h]
\vspace{-1cm}
\begin{centering}
\includegraphics[scale=0.6]{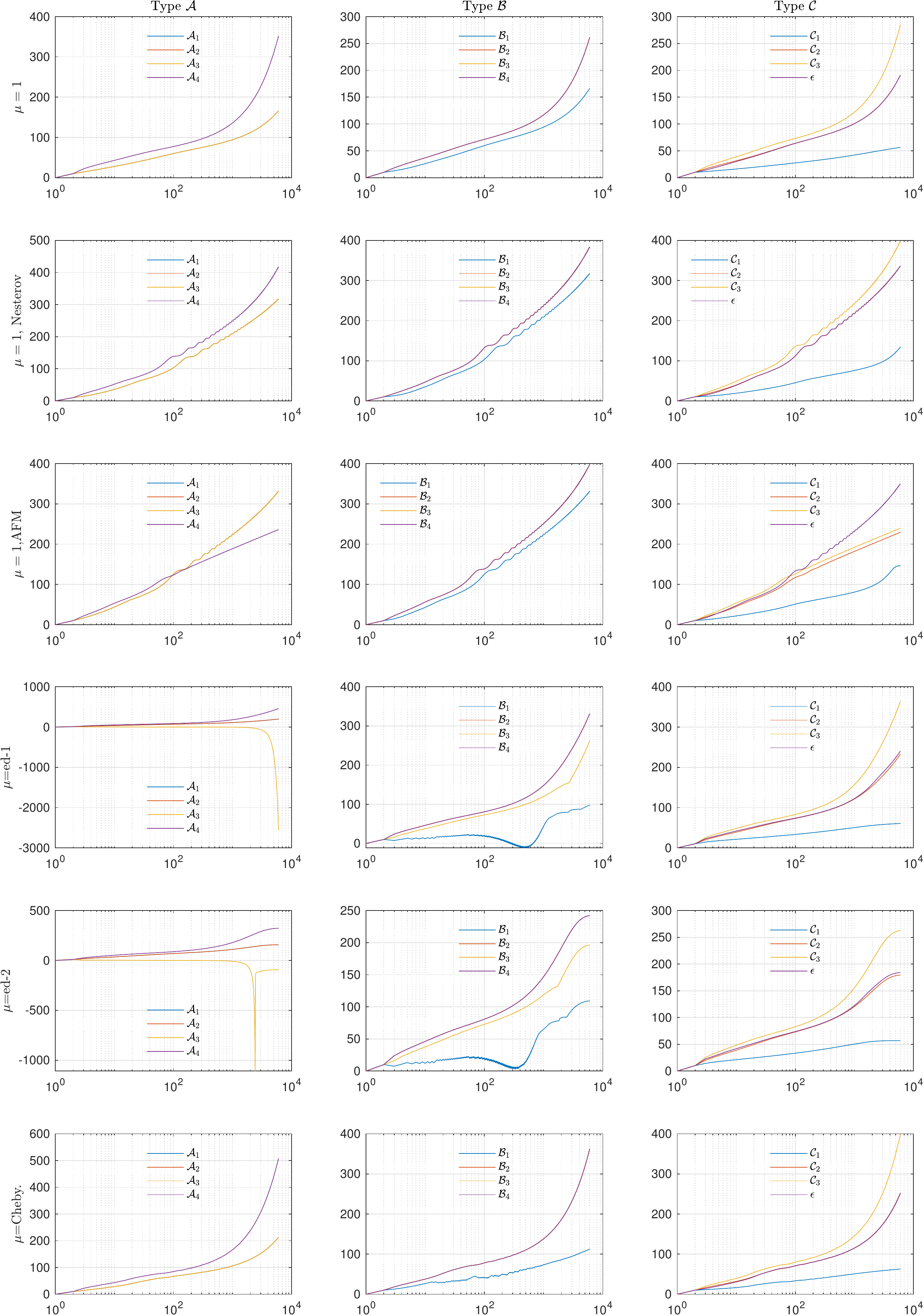}
\par\end{centering}
\caption{\label{fig:Results-of-Gaussian1}Results of Gaussian filter with $\sigma=1$.}
\end{figure}

\begin{figure}[h]
\vspace{-1cm}
\begin{centering}
\includegraphics[scale=0.6]{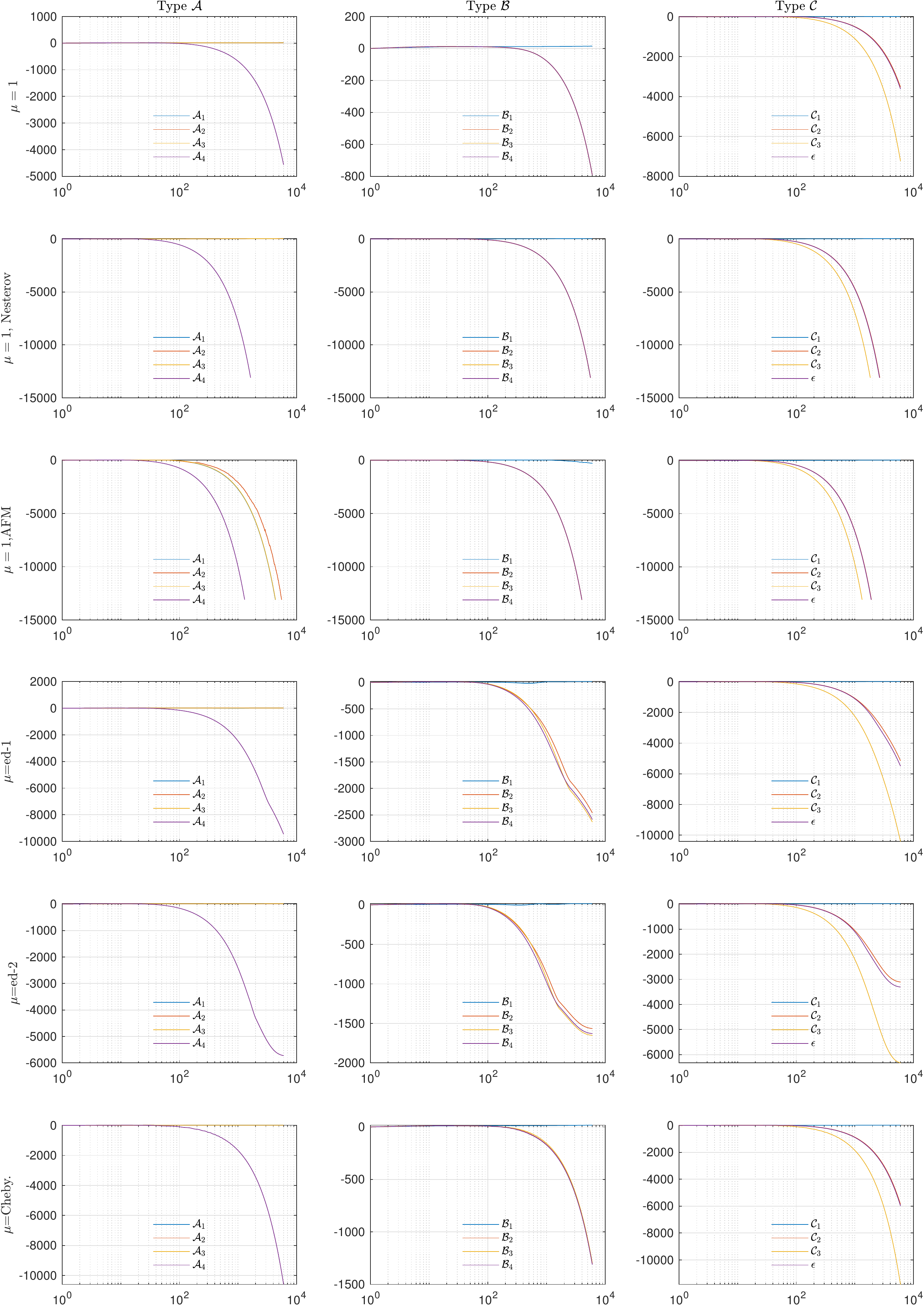}
\par\end{centering}
\caption{\label{fig:Results-of-Gaussian5}Results of Gaussian filter with $\sigma=5$.}
\end{figure}

\begin{figure}[h]
\vspace{-1cm}
\begin{centering}
\includegraphics[scale=0.6]{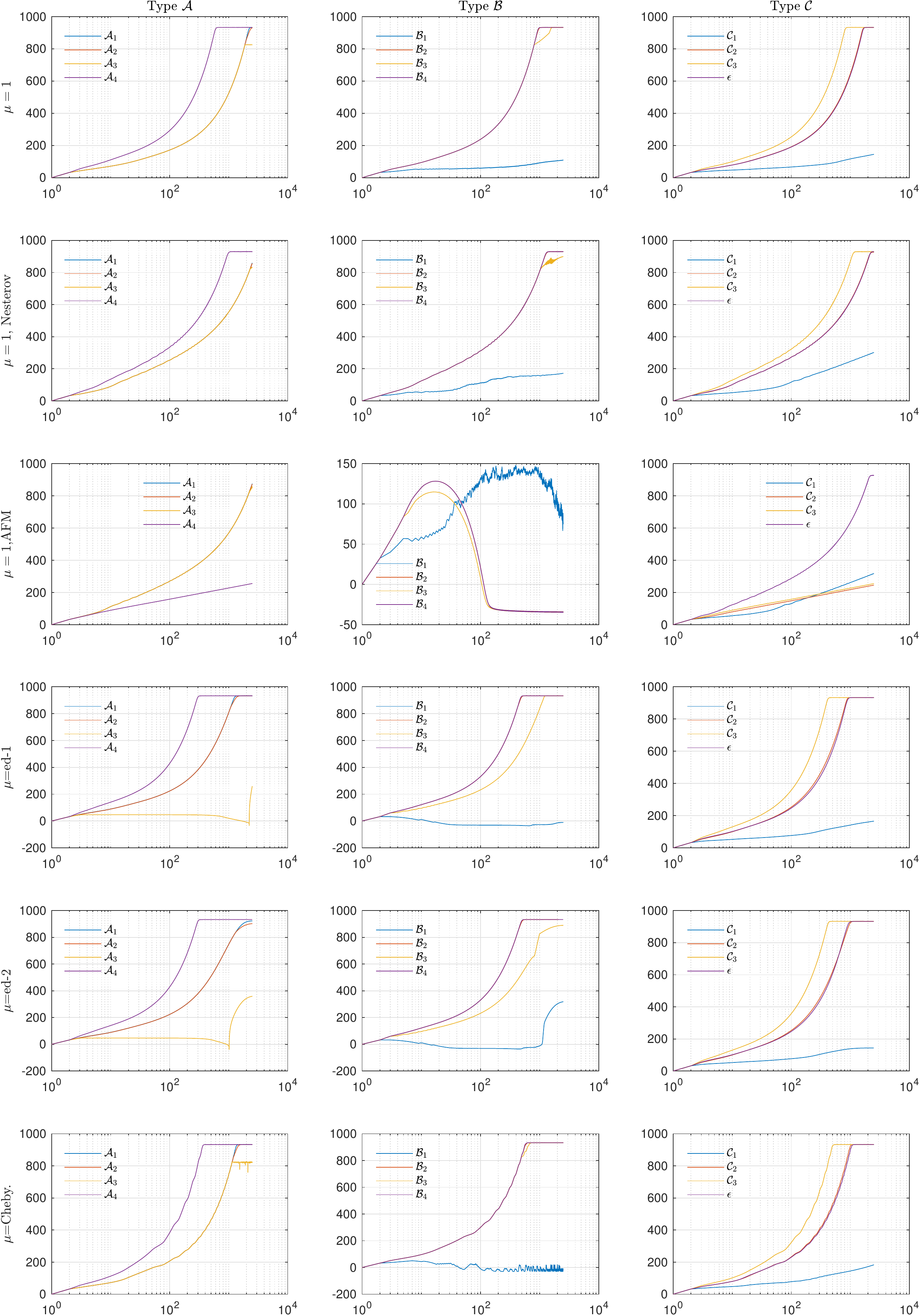}
\par\end{centering}
\caption{\label{fig:Results-of-guided}Results of guided filter.}
\end{figure}

\begin{figure}[h]
\vspace{-1cm}
\begin{centering}
\includegraphics[scale=0.6]{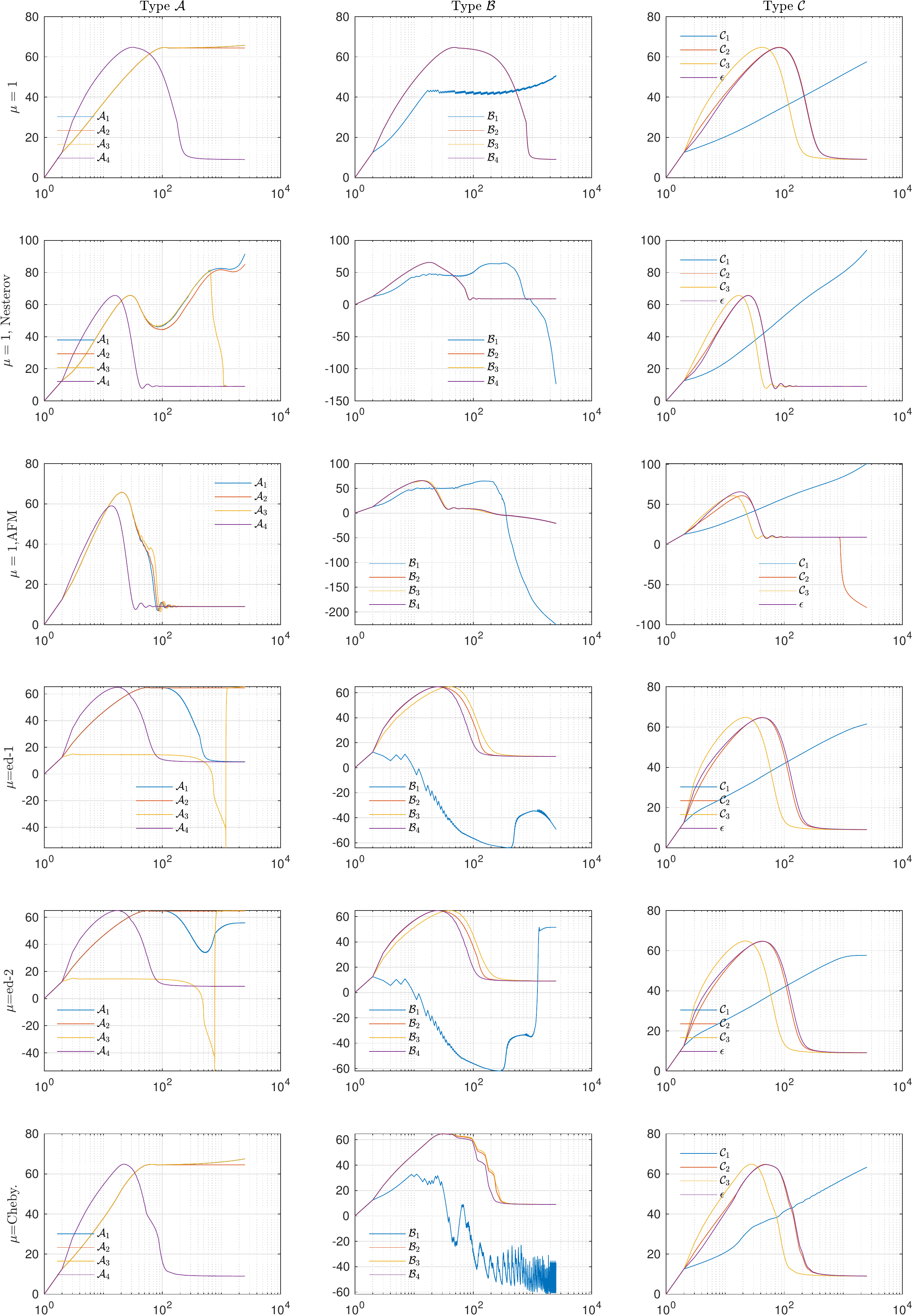}
\par\end{centering}
\caption{\label{fig:Results-of-bilateral}Results of bilateral filter.}
\end{figure}

Results for the parametric-free methods and parametric methods are
presented in Figures \ref{fig:Results-of-Gaussian1} to \ref{fig:Results-of-bilateral}.
Key observations are summarized as follows.
\begin{enumerate}
\item Parametric-free methods (first 3 rows of Figures \ref{fig:Results-of-Gaussian1}
to \ref{fig:Results-of-bilateral})
\begin{enumerate}
\item Gaussian filter with $\sigma=1$.\\

\begin{enumerate}
\item Without acceleration, it is shown in the first row of the figure that
the top performers in each group are $\mathcal{{A}}_{4}$, $\mathcal{{B}}_{2-4}$,
and $\mathcal{{C}}_{3}$. Among them $\mathcal{{A}}_{4}$ is better
in terms of achieving the highest PSNR improvement.
\item Results shown in the 2nd and 3rd row of the figure demonstrate that
the two acceleration methods generally improve the speed of PSNR improvement
for all methods.
\item The Nesterov acceleration performs better than the AFM method in that
it leads to better PSNR improvement at the end of the iteration for
Types $\mathcal{A}$ and $\mathcal{C}$. For Type $\mathcal{B}$,
the PSNR improvements are about the same for the two methods.
\end{enumerate}
\item Gaussian filter with $\sigma=5$.
\begin{enumerate}
\item Only 5 methods results in improvement of PSNR:$\mathcal{{A}}_{1}$,
$\mathcal{{A}}_{2},\mathcal{{A}}_{3},\mathcal{{B}}_{1}$, and $\mathcal{{C}}_{1}$.
The improvement in PSNR is small (about 15\%).
\item Nesterov acceleration helps improve the speed of PSNR improvement
for all convergent methods (increase from 15\% to about 40\%). The
AFM acceleration leads to divergent results for all methods, except
for $\mathcal{{C}}_{1}$.
\end{enumerate}
\item Guided filter.
\begin{enumerate}
\item Without acceleration, the top performers in each group are $\mathcal{{A}}_{1,2,4}$,
$\mathcal{{B}}_{2-4}$, and $\mathcal{{C}}_{3}$ and $\epsilon$.
Among them $\mathcal{{A}}_{4}$ $\mathcal{{B}}_{4}$, and $\mathcal{{C}}_{3}$
are better in terms of speed of improvement.
\item The Nesterov acceleration generally improves the speed for all methods.
Results of using the AFM acceleration are mixed. While methods such
as $\mathcal{{A}}_{3}$ and $\epsilon$ benefit from the acceleration,
other methods such as the group $\mathcal{{B}}$ methods become divergent.
\item The acceleration in convergence is at the cost of converging to a
slightly lower improvement in PSNR than that without acceleration.
For example, without acceleration the $\mathcal{A}_{1}$ method converges
to 932\% PSNR improvement. With Nesterov or AFM, the same method converges
to 857\% and 875\%, respectively. We should point out that such differences
are of no practical importance in this application, because the three
PSNR improvements correspond to PSNR values of 295dB, 274dB and 279dB,
respectively. The reverse filtering results in these three cases are
effectively the same as the original image.
\end{enumerate}
\item Bilateral filter.
\begin{enumerate}
\item Only 5 methods lead to PSNR improvement:$\mathcal{{A}}_{1}$, $\mathcal{{A}}_{2},\mathcal{{A}}_{3},\mathcal{{B}}_{1}$,
and $\mathcal{{C}}_{1}$. The improvement in PSNR is modest (about
60\%).
\item Both acceleration methods fail to improve the speed. The exception
is method $\mathcal{{C}}_{1}$ for which the two acceleration methods
lead to the faster speed of improvement.
\end{enumerate}
\item Summary. Among the tested methods, $\mathcal{{A}}{4}$, $\mathcal{{B}}{4}$,
and $\mathcal{{C}}{3}$ were found to be the most effective for reversing
the effects of two filters, namely Gaussian with $\sigma=1$ and guided
filter, which are relatively easy to reverse. On the other hand, for
two filters, namely Gaussian with $\sigma=5$ and bilateral filter,
which are more difficult to reverse, the top-performing methods were
$\mathcal{{A}}{1}$, $\mathcal{{B}}{1}$, and $\mathcal{{C}}{1}$.
Such observations will be used in the main paper as the basis to compare
different vector Steffensen's methods and their variants.
\end{enumerate}
\item Parametric methods (row 4 to row 6 in Figures \ref{fig:Results-of-Gaussian1}
to \ref{fig:Results-of-bilateral}).\\
In general, the list of methods, that result in PSNR improvement,
is the same as that in the parametric-free case.
\begin{enumerate}
\item Gaussian filter with $\sigma=1$.
\begin{enumerate}
\item The methods ``ed-1'' and ``Cheby.'' appear to produce comparable
or superior performance to parametric-free methods in terms of speed
of PSNR improvement. For example, method $\mathcal{A}_{4}$ with $\mu=$ed-1
and $\mu=$Cheby. achieves 460\% and 508\% PSNR improvement, respectively.
In contrast, the same method with $\mu=1$ (parameter-free setting)
achieves 352\%.
\item The method ``ed-2'' is inferior to ``ed-1'' and ``Cheby.'' and
its performance is not as good as that the parameter-free case.
\item In general, the parametric-free methods with Nesterov or AFM achieve
better PSNR improvement than those corresponding parametric methods.
\end{enumerate}
\item Gaussian filter with $\sigma=5$.
\begin{enumerate}
\item The methods ``ed-1'', ``ed-2'' and ``Cheby.'' appear to produce
comparable performance to parametric-free methods in terms of speed
of PSNR improvement. There is no benefit of using these method.
\end{enumerate}
\item Guided filter
\begin{enumerate}
\item The methods ``ed-1'', ``ed-2'' and ``Cheby.'' appear to produce
comparable performance to parametric-free methods in terms of speed
PSNR improvement.
\item There are a few exceptions. (1) The $\mathcal{A}_{3}$ method with
``ed-1'' and ``ed-2'' lead to far worse result than the corresponding
method without using the parameter. (2) while ``ed-1'' and ``Cheby.''
make the method $\mathcal{{B}}_{1}$ divergent, ``ed-2'' significantly
improves the PSNR improvement of $\mathcal{{B}}_{1}$ from the parameter-free
case of 109\% to 318\%. (3) ``ed-2'' makes $\mathcal{{B}}_{2}$
divergent.
\end{enumerate}
\item Bilateral filter
\begin{enumerate}
\item The methods ``ed-1'', ``ed-2'' and ``Cheby.'' appear to produce
comparable performance to parametric-free methods in terms of speed
of PSNR improvement.
\item The exception is that both ``ed-1'' and ``Cheby.'' make the method
$\mathcal{{B}}_{1}$ divergent.
\end{enumerate}
\end{enumerate}
\end{enumerate}

\subsubsection*{Parametric methods with acceleration methods of Nesterov and AFM
\label{subsec:Parametric-methods-with}}

\begin{figure}[h]
\vspace{-1cm}
\begin{centering}
\includegraphics[scale=0.6]{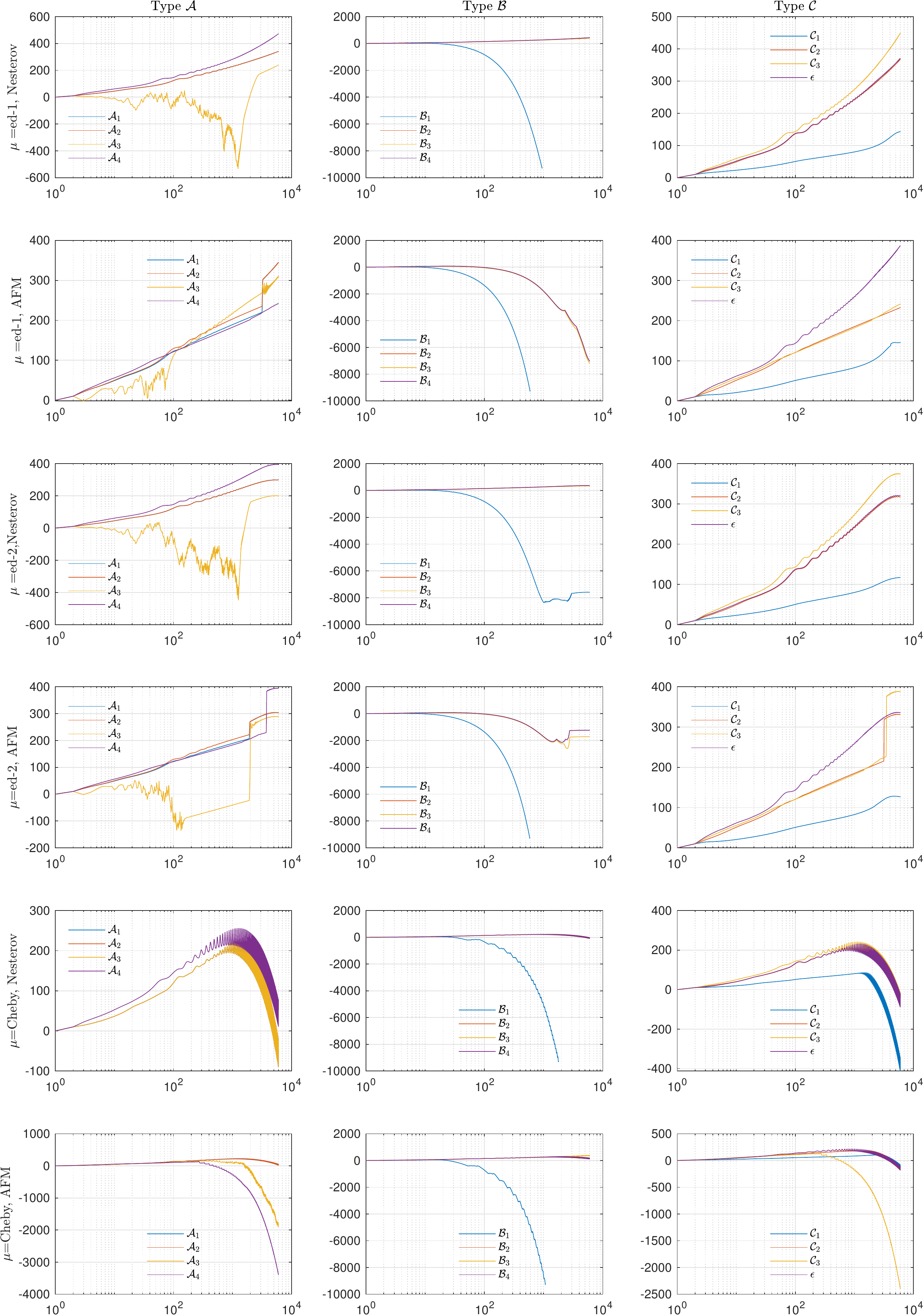}
\par\end{centering}
\caption{\label{fig:Results-of-Gaussian1-1}Results of Gaussian filter with
$\sigma=1$.}
\end{figure}

\begin{figure}[h]
\vspace{-1cm}
\begin{centering}
\includegraphics[scale=0.6]{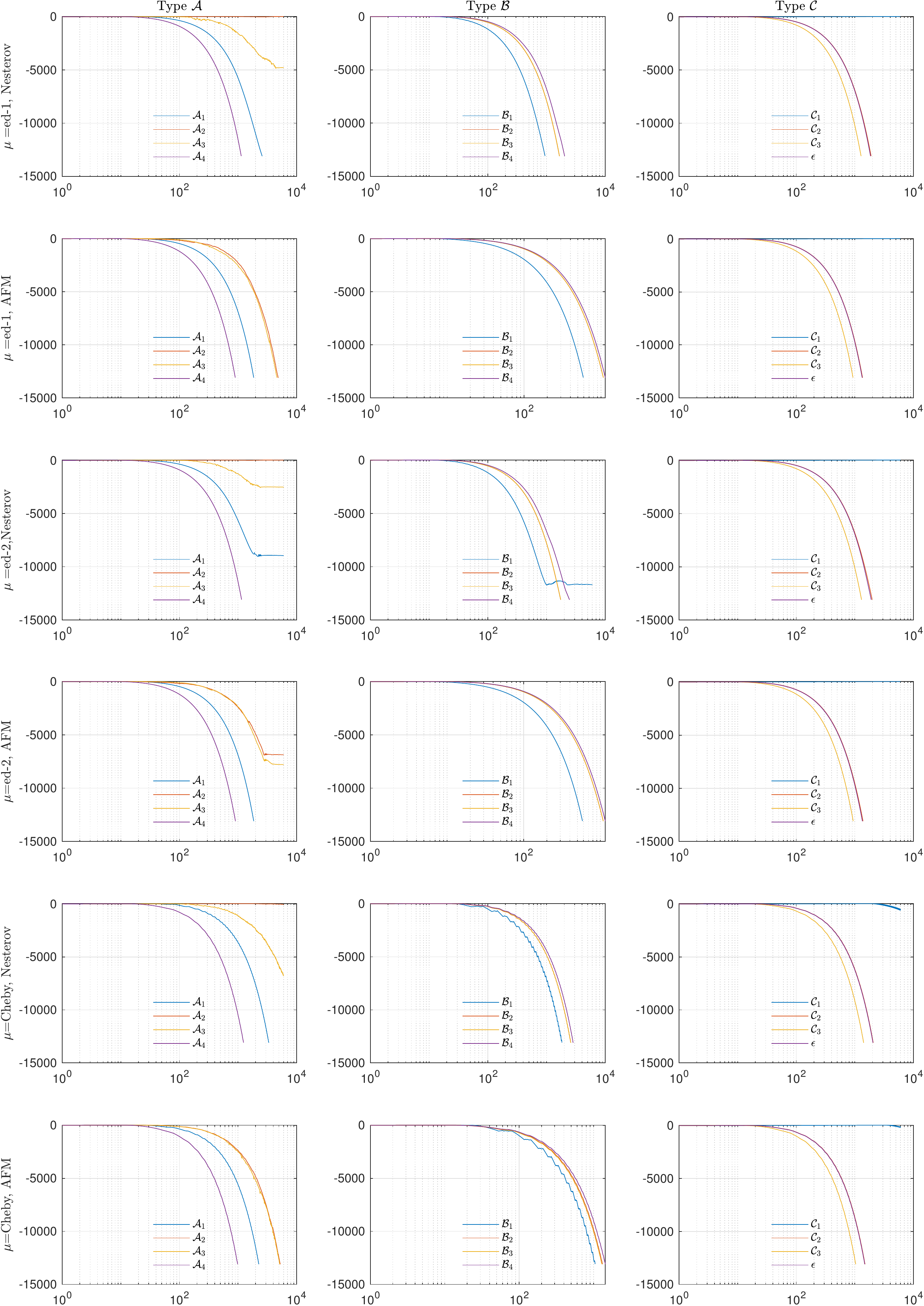}
\par\end{centering}
\caption{\label{fig:Results-of-Gaussian5-1}Results of Gaussian filter with
$\sigma=5$.}
\end{figure}

\begin{figure}[h]
\vspace{-1cm}
\begin{centering}
\includegraphics[scale=0.6]{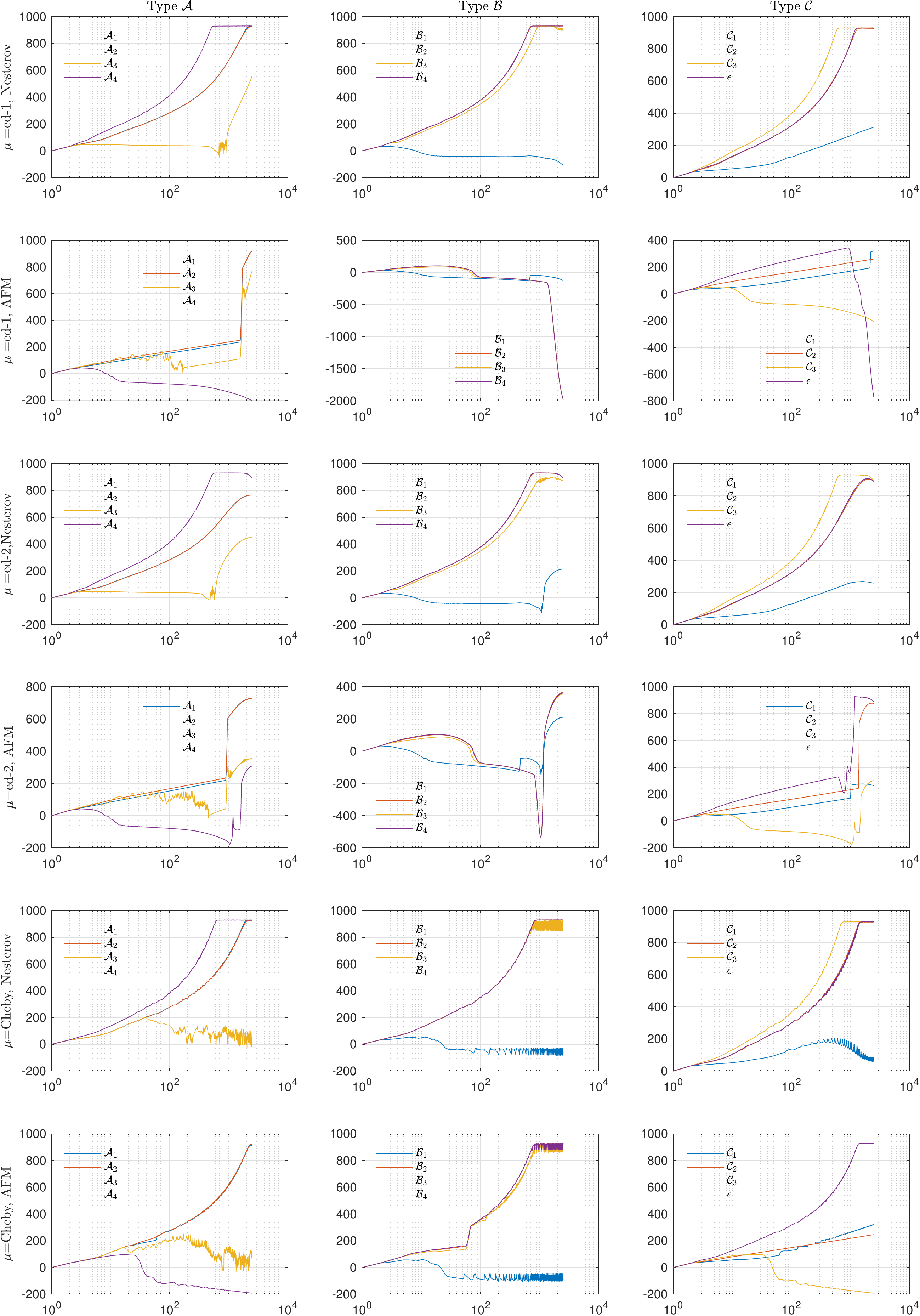}
\par\end{centering}
\caption{\label{fig:Results-of-guided-1}Results of guided filter.}
\end{figure}

\begin{figure}[h]
\vspace{-1cm}
\begin{centering}
\includegraphics[scale=0.6]{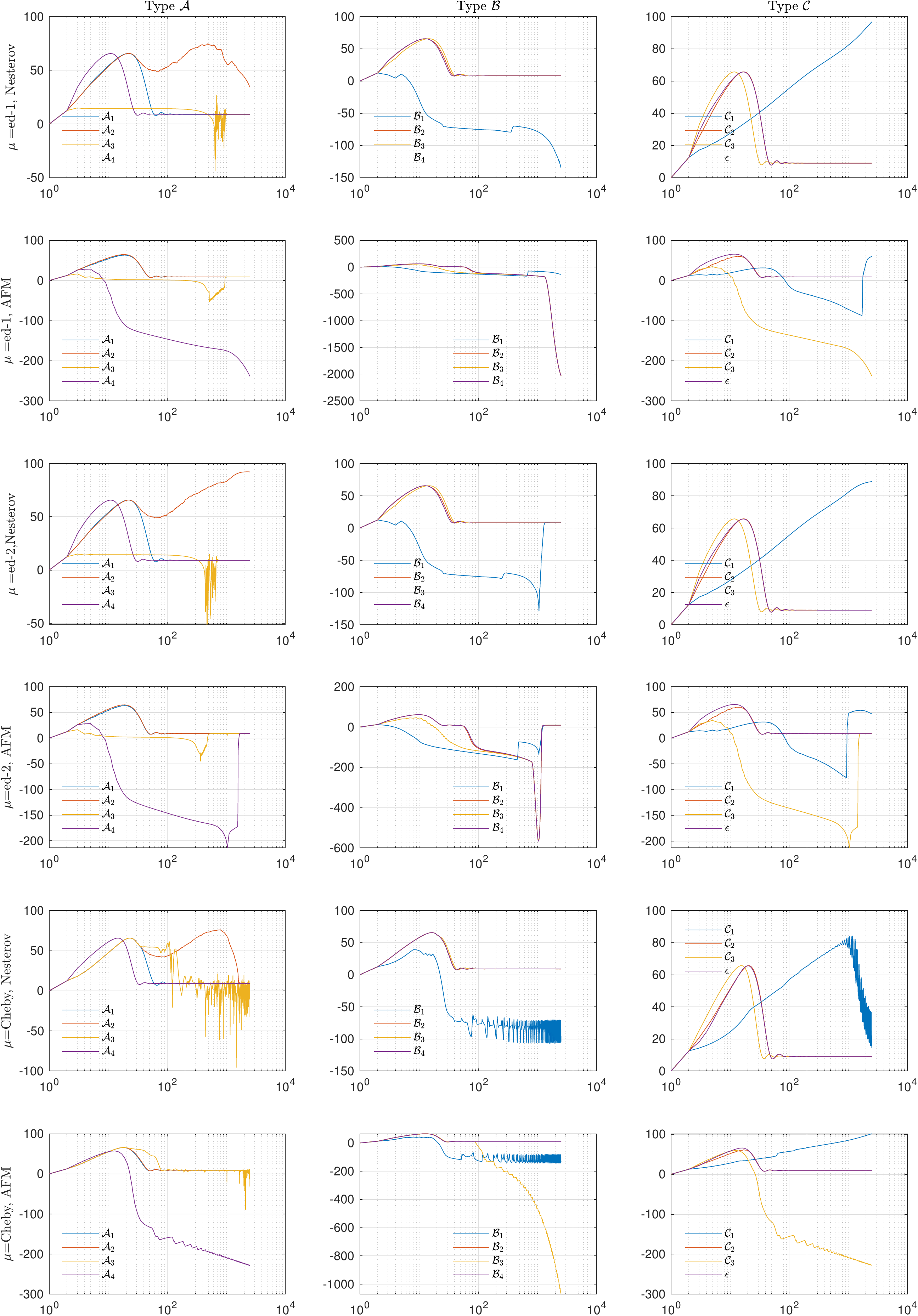}
\par\end{centering}
\caption{\label{fig:Results-of-bilateral-1}Results of bilateral filter.}
\end{figure}

Results for the accelerated parametric methods are presented in Figures
\ref{fig:Results-of-Gaussian1-1} to \ref{fig:Results-of-bilateral-1}.
Comparing the two groups of figures (Figures \ref{fig:Results-of-Gaussian1}-\ref{fig:Results-of-bilateral}
and Figures\ref{fig:Results-of-Gaussian1-1}-\ref{fig:Results-of-bilateral-1})
we have the following observations.
\begin{enumerate}
\item Gaussian filter with $\sigma=1$.
\begin{enumerate}
\item The Type $\mathcal{A}$ and $\mathcal{C}$ methods with \textquotedblleft ed-1\textquotedblright{}
or \textquotedblleft ed-2\textquotedblright{} settings and Nesterov
or AFM acceleration outperform or match performance of the parametric-free
methods in terms of PSNR improvement at the end of the iteration.
However, the ``Cheby'' setting with Nesterov or AFM acceleration
leads to divergence or worse results.
\item For the Type $\mathcal{B}$ methods, only the setting of ``ed-1''
together with Nesterov acceleration have produces about the same or
better results than parametric-free methods in terms of PSNR improvement
at the end of the iteration. Other combinations of parameter setting
and acceleration lead to either inferior or divergent results.
\end{enumerate}
\item Gaussian filter with $\sigma=5$.
\begin{enumerate}
\item The settings ``ed-1'' or ``ed-2'' together with Nesterov or AFM
acceleration have produced comparable performance to parametric-free
methods in terms of PSNR improvement at the end of the iteration.
\item Other combinations of parameter setting and acceleration lead to either
inferior or divergent results.
\end{enumerate}
\item Guided filter
\begin{enumerate}
\item Only the two methods $\mathcal{{B}}_{1}$ and $\mathcal{{C}}_{1}$
have benefited from combinations of one of the three parameter settings
together with the acceleration methods (Nesterov and AFM). Using such
combinations, all other methods have produced comparable or inferior
results. As such, there is no benefit of using further accelerations
in this case.
\end{enumerate}
\item Bilateral filter
\begin{enumerate}
\item The methods ``ed-1'', ``ed-2'' and ``Cheby.'' appear to produce
comparable performance to parametric-free methods in terms of speed
of convergence and PSNR improvement. The exception is that both ``ed-1''
and ``Cheby.'' make the method $\mathcal{{B}}_{1}$ divergent.
\end{enumerate}
\end{enumerate}

\subsection*{Performance of existing methods\label{subsec:Performance-of-existing}}

Results for the 4 existing methods are presented in Fig. \ref{fig:three graphs}
and Table \ref{tab:subtables}. We can summarize results as follows.
\begin{enumerate}
\item Gaussian filter with $\sigma=1$.
\begin{enumerate}
\item The S-method is a clear winner in terms of achieving the highest PSNR
improvement.
\item The two acceleration methods improve the convergence speed of the
T-method, but the impact on other methods is mixed. Specifically,
the Nesterov acceleration can cause the P-method and S-method to diverge,
while the AFM method cause the S-method to diverge.
\end{enumerate}
\item Gaussian filter with $\sigma=5$.
\begin{enumerate}
\item Only the TDA-method and P-method have shown PSNR improvement as the
iteration progresses. The T-method and the S-method fail to improve
the PSNR.
\item The two acceleration methods substantially improves the speed of PSNR
improvement of the TDA-method and the P-method.
\item The TDA-method is the overall winner due to is low computational complexity
compared to that of the P-method.
\end{enumerate}
\item Guided filter
\begin{enumerate}
\item The T-method and S-method have achieved about the same PSNR improvement.
However, the T-method is preferred because of its low computational
complexity.
\item The Nesterov acceleration only improves the performance of the TDA
method, and it causes other methods to produce inferior results compared
to those without the acceleration.
\item The AFM acceleration have a negative impact on all four methods.
\end{enumerate}
\item Bilateral filter
\begin{enumerate}
\item Both the T-method and S-method exhibit PSNR improvements that peaks
at 9\%. In contrast, the TDA-method and P-method show a PSNR improvement
peak at around 97\%.
\item The Nesterov acceleration has a positive effect on the speed of PSNR
improvement, while the AFM acceleration has a negative impact on the
T, P, and TDA methods and has only a minor effect on the S-method.
\end{enumerate}
\end{enumerate}

\section*{Best performing methods for inverting effects of the four filters\label{subsec:Convergence-and-comparison}}

From the above results, the best performing methods for reversing
effects of each filter are listed in the following Table \ref{tab:Best-performing-methods.}.
A detailed discussion of results due to these methods are presented
in the main paper.
\begin{table*}[t]
\begin{centering}
\begin{tabular}{|c|c|c|c||c|c|}
\hline 
 & \multicolumn{3}{c||}{Vector variable Steffensen's method} & \multicolumn{2}{c|}{T, TDA, P, S}\tabularnewline
\hline 
\hline 
Gauss-1  & $\mathcal{{A}}_{4},\mu=$Cheby., (508) & $\mathcal{{B}}_{4},\mu=$ed-1, Nest., (430) & $\mathcal{{C}}_{3},\mu=$ed-1, Nest., (449) & T, AFM (350) & S (717)\tabularnewline
\hline 
Gauss-5 & $\mathcal{{A}}_{1},$ Nest., (40) & $\mathcal{{B}}_{1},$Nest., (40) & $\mathcal{{C}}_{1},\mu=$ed-1, AFM., (45) & TDA, AFM, (45) & P, AFM, (45)\tabularnewline
\hline 
Guided & $\mathcal{{A}}_{4}$ (933) & $\mathcal{{B}}_{4}$ (933) & $\mathcal{{C}}_{3}$ (933) & T (921) & S (920)\tabularnewline
\hline 
Bilateral & $\mathcal{{A}}_{1},$Nest., (91) & $\mathcal{{B}}_{1},\mu=$ed-2, (52) & $\mathcal{{C}}_{1},$AFM, (101) & TDA, Nest., (97) & P, Nest., (98)\tabularnewline
\hline 
\end{tabular}
\par\end{centering}
\caption{\label{tab:Best-performing-methods.}Best performing methods chosen
based solely on the highest percentage of PSNR improvement when the
method stops at the maximum number of preset iterations. Methods without
specification of $\mu_{n}$ are with the setting $\mu_{n}=1$. The
number in bracket, e.g., (508), indicates the 508\% PSNR improvement
at the end of the iteration. Nesterov acceleration is abbreviated
as Nest..}
\end{table*}

\newpage

\begin{table}                                                 
\centering
\begin{tabular}{|c|c|c|c|c|c|c|c|c|c|c|c|c|}                                                                             
\hline
 & $\mathcal{A}_1$ & $\mathcal{A}_2$ & $\mathcal{A}_3$ & $\mathcal{A}_4$ & $\mathcal{B}_1$ & $\mathcal{B}_2$ & $\mathcal{B}_3$ & $\mathcal{B}_4$ & $\mathcal{C}_1$ & $\mathcal{C}_2$ & $\mathcal{C}_3$ & $\epsilon$ \\
\hline            
$\mu=1$ & 166 & 166 & 166 & 352 & 166 & 261 & 261 & 261 & 56 & 191 & 284 & 191 \\                                        
\hline    
$\mu=1$, Nest. & 317 & 317 & 317 & 417 & 317 & 383 & 383 & 383 & 134 & 336 & 398 & 336 \\                                
\hline                                                                                       
$\mu=1$, AFM & 331 & 331 & 331 & 236 & 331 & 398 & 398 & 398 & 146 & 229 & 240 & 349 \\                                 
\hline                                                                                     
$\mu$=ed1 & 199 & 199 & - & 460 & 98 & 331 & 262 & 331 & 61 & 233 & 364 & 240 \\       
\hline
$\mu=$ed2 & 157 & 157 & - & 323 & 109 & 242 & 196 & 242 & 57 & 179 & 263 & 184 \\  
\hline
$\mu$=Cheby & 213 & 213 & 213 & 508 & 112 & 362 & 362 & 362 & 63 & 251 & 399 & 252 \\      
\hline
\hline                                                                                                                   
ed1+Nest. & 341 & 341 & 240 & 472 & - & 430 & 388 & 430 & 143 & 367 & 449 & 370 \\                                       
\hline                                                                                                                   
ed1+AFM & 345 & 345 & 310 & 242 & - & - & - & - & 146 & 232 & 242 & 387 \\
\hline                                                                                                                   
ed2+Nest. & 299 & 298 & 200 & 394 & - & 360 & 336 & 360 & 117 & 317 & 375 & 320 \\
\hline                                                                                                                   
ed2+AFM & 304 & 304 & 290 & 394 & - & - & - & - & 127 & 331 & 388 & 336 \\ 
\hline                                                                                                                   
Cheby.+Nest. & - & - & - & 57 & - & - & - & - & - & - & - & - \\
\hline                                                                                                                   
Cheby.+AFM & 40 & 43 & - & - & - & 144 & 354 & 133 & - & - & - & - \\  
\hline                             
\end{tabular}    
\bigskip
\caption{Percentage of PSNR improvement at the end of iteration for the Gaussian filter $\sigma=1$. The entry with "-" means the method is divergent.}                    
\label{tab:Gauss1}
\end{table}          

\begin{table}
\centering
\begin{tabular}{|c|c|c|c|c|c|c|c|c|c|c|c|c|}
\hline
 & $\mathcal{A}_1$ & $\mathcal{A}_2$ & $\mathcal{A}_3$ & $\mathcal{A}_4$ & $\mathcal{B}_1$ & $\mathcal{B}_2$ & $\mathcal{B}_3$ & $\mathcal{B}_4$ & $\mathcal{C}_1$ & $\mathcal{C}_2$ & $\mathcal{C}_3$ & $\epsilon$ \\
\hline                                                                                                                   
$\mu=1$ & 14 & 12 & 14 & - & 14 & - & - & - & 16 & - & - & - \\        
\hline                                                                                                                   
$\mu=1$, Nest. & 40 & 39 & 40 & - & 40 & - & - & - & 41 & - & - & - \\                                                   
\hline                                                                                                                   
$\mu=1$, AFM & - & - & - & - & - & - & - & - & 44 & - & - & - \\ 
\hline                                                                                                                   
$\mu$=ed1 & 16 & 12 & 12 & - & 16 & - & - & - & 19 & - & - & - \\  
\hline                                                                                                                   
$\mu=$ed2 & 15 & 12 & 13 & - & 15 & - & - & - & 17 & - & - & - \\    
\hline                                                                                                                   
$\mu$=Cheby & 17 & 12 & 17 & - & 17 & - & - & - & 20 & - & - & - \\    
\hline
\hline
ed1+Nest. & - & 25 & - & - & - & - & - & - & 43 & - & - & - \\
\hline                                                                                                                   
ed1+AFM & - & - & - & - & - & - & - & - & 45 & - & - & - \\                                                              
\hline                                                                                                                   
ed2+Nest. & - & 21 & - & - & - & - & - & - & 40 & - & - & - \\   
\hline                                                                                                                   
ed2+AFM & - & - & - & - & - & - & - & - & 42 & - & - & - \\    
\hline                                                                                                                   
Cheby.+Nest. & - & - & - & - & - & - & - & - & - & - & - & - \\  
\hline                                                                                                                   
Cheby.+AFM & - & - & - & - & - & - & - & - & - & - & - & - \\ 
\hline             
\end{tabular}
\bigskip
\caption{Percentage of PSNR improvement at the end of iteration for the Gaussian filter  $\sigma=5$. The entry with "-" means the method is divergent}                     
\label{table:Gauss5}                                                               
\end{table}        

\clearpage

\begin{table}                                                                              
\centering                                                                                
\begin{tabular}{|c|c|c|c|c|c|c|c|c|c|c|c|c|}                                                 
\hline                                                 
 & $\mathcal{A}_1$ & $\mathcal{A}_2$ & $\mathcal{A}_3$ & $\mathcal{A}_4$ & $\mathcal{B}_1$ & $\mathcal{B}_2$ & $\mathcal{B}_3$ & $\mathcal{B}_4$ & $\mathcal{C}_1$ & $\mathcal{C}_2$ & $\mathcal{C}_3$ & $\epsilon$ \\
\hline                                                                                       
$\mu=1$ & 932 & 930 & 825 & 933 & 109 & 933 & 933 & 933 & 145 & 933 & 933 & 933 \\                                       
\hline                                                                                                                   
$\mu=1$, Nest. & 857 & 855 & 833 & 930 & 172 & 931 & 899 & 930 & 301 & 928 & 930 & 929 \\                                
\hline                                                                                                                   
$\mu=1$, AFM & 875 & 872 & 851 & 255 & 75 & - & - & - & 318 & 245 & 255 & 927 \\ 
\hline                                                                                                                   
$\mu$=ed1 & 933 & 933 & 258 & 933 & - & 933 & 933 & 933 & 166 & 933 & 933 & 933 \\    
\hline                                      
$\mu=$ed2 & 923 & 902 & 359 & 933 & 318 & - & 890 & 933 & 144 & 933 & 933 & 932 \\                                       
\hline
$\mu$=Cheby & 933 & 933 & 825 & 933 & - & 933 & 933 & 933 & 183 & 933 & 933 & 933 \\
\hline  
\hline                                                                                                                   
ed1+Nest. & 927 & 927 & 558 & 930 & - & 930 & 902 & 930 & 313 & 927 & 930 & 929 \\
\hline                                                                                                                   
ed1+AFM & 921 & 919 & 772 & - & - & - & - & - & 322 & 261 & - & - \\ 
\hline                                                                                                                   
ed2+Nest. & 766 & 766 & 448 & 893 & 214 & 893 & 872 & 893 & 258 & 887 & 893 & 890 \\                                     
\hline                                                                                                                   
ed2+AFM & 727 & 726 & 352 & 308 & 210 & 367 & 356 & 363 & 264 & 875 & 301 & 889 \\                                       
\hline                                                                                                                   
Cheby.+Nest. & 927 & 927 & 26 & 930 & - & 929 & 844 & 930 & 59 & 928 & 930 & 929 \\         
\hline                                                                                                                   
Cheby.+AFM & 924 & 923 & 78 & - & - & 927 & 921 & 927 & 322 & 246 & - & 927 \\ 
\hline                           
\end{tabular}      
\bigskip
\caption{Percentage of PSNR improvement at the end of iteration for the guided filter. The entry with "-" means the method is divergent.}                                   
\label{table:Guided_Table}                                                               
\end{table}  
\vspace{3cm}                 
\begin{table}[h!]              
\centering                  
\begin{tabular}{|c|c|c|c|c|c|c|c|c|c|c|c|c|}
\hline                        
& $\mathcal{A}_1$ & $\mathcal{A}_2$ & $\mathcal{A}_3$ & $\mathcal{A}_4$ & $\mathcal{B}_1$ & $\mathcal{B}_2$ & $\mathcal{B}_3$ & $\mathcal{B}_4$ & $\mathcal{C}_1$ & $\mathcal{C}_2$ & $\mathcal{C}_3$ & $\epsilon$ \\
\hline
\hline    
\hline                                                                                       
$\mu=1$ & 66 & 64 & 66 & 9 & 50 & 9 & 9 & 9 & 58 & 9 & 9 & 9 \\                                                          
\hline                                                                                                                   
$\mu=1$, Nest. & 91 & 85 & 9 & 9 & - & 9 & 9 & 9 & 94 & 9 & 9 & 9 \\                                                 
\hline                                                                                                                   
$\mu=1$, AFM & 9 & 9 & 9 & 9 & - & - & - & - & 101 & - & 9 & 9 \\                                            
\hline                                                                                                                   
$\mu$=ed1 & 9 & 64 & 65 & 9 & - & 9 & 9 & 9 & 62 & 9 & 9 & 9 \\                                                      
\hline                                                                                                                   
$\mu=$ed2 & 56 & 65 & 65 & 9 & 52 & 9 & 9 & 9 & 58 & 9 & 9 & 9 \\                                                        
\hline                                                                                                                  
$\mu$=Cheby & 67 & 64 & 67 & 9 & - & 9 & 9 & 9 & 63 & 9 & 9 & 9 \\                
\hline
\hline    
ed1+Nest. & 9 & 34 & 9 & 9 & - & 9 & 9 & 9 & 97 & 9 & 9 & 9 \\                                                       
\hline                                                                                                                   
ed1+AFM & 9 & 9 & 9 & - & - & - & - & - & 60 & 9 & - & 9 \\                                        
\hline                                                                                                                   
ed2+Nest. & 9 & 92 & 9 & 9 & 9 & 9 & 9 & 9 & 89 & 9 & 9 & 9 \\                                                           
\hline                                                                                                                   
ed2+AFM & 9 & 9 & 9 & 9 & 9 & 9 & 9 & 9 & 47 & 9 & 9 & 9 \\                                                              
\hline                                                                                                                   
Cheby.+Nest. & 9 & 9 & 13 & 9 & - & 9 & 9 & 9 & 14 & 9 & 9 & 9 \\                                                    
\hline                                                                                                                   
Cheby.+AFM & 9 & 9 & -3 & - & - & 9 & - & 9 & 101 & 9 & - & 9 \\      
\hline                                                                             
\end{tabular}         
\bigskip
\caption{Percentage of PSNR improvement at the end of iteration for the bilateral filter. The entry with "-" means the method is divergent.}
\label{tab:MyTableLabel-4}
\end{table}              

\begin{figure}[h!]
     \centering
     \begin{subfigure}[b]{0.49\textwidth}
         \centering
         \includegraphics[width=\textwidth,viewport=35bp 25bp 690bp 730bp,clip]{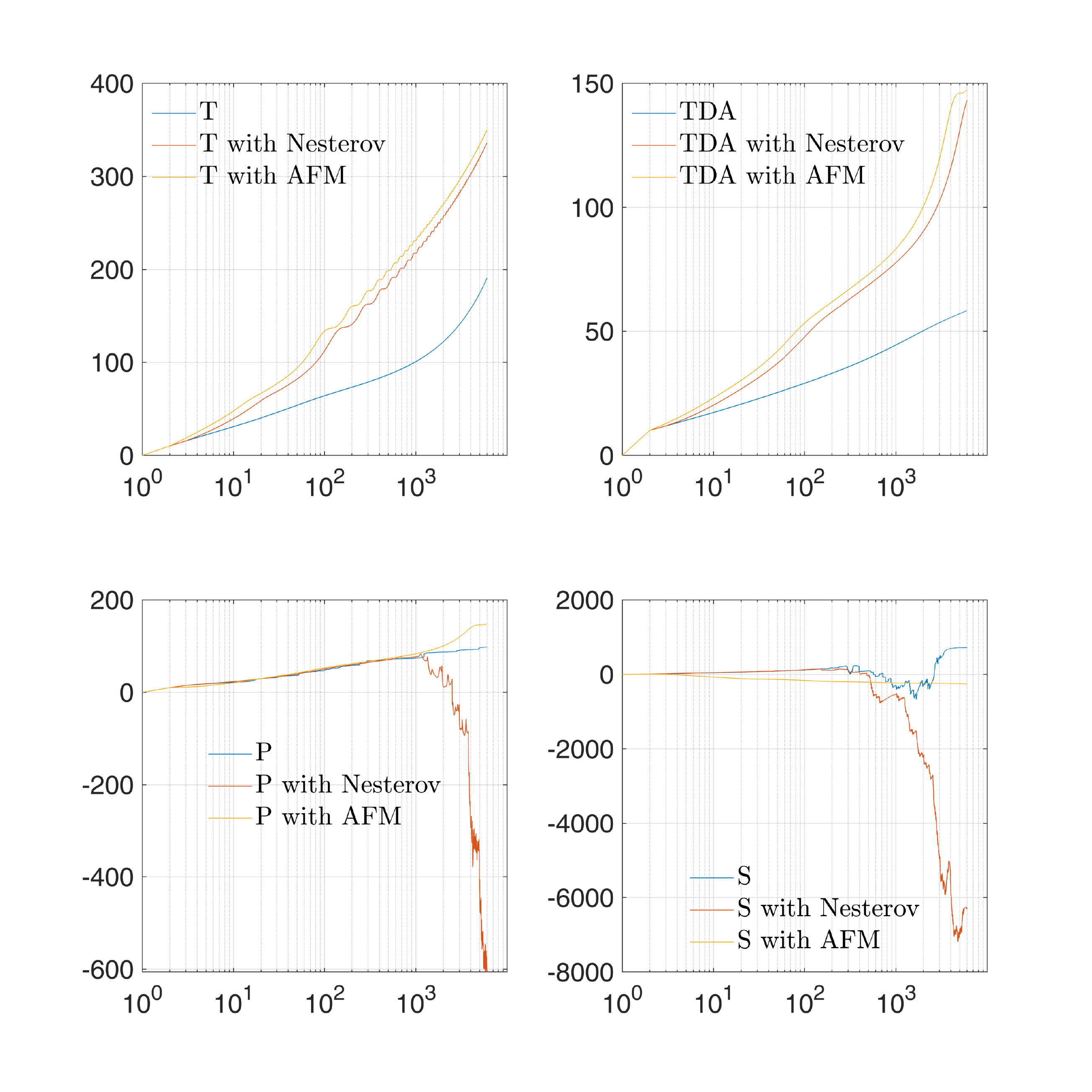}
         \caption{Gaussian ($\sigma=1$)}
     \end{subfigure}
     \hspace{1mm}
     \begin{subfigure}[b]{0.49\textwidth}
         \centering
         \includegraphics[width=\textwidth,viewport=35bp 25bp 690bp 730bp,clip]{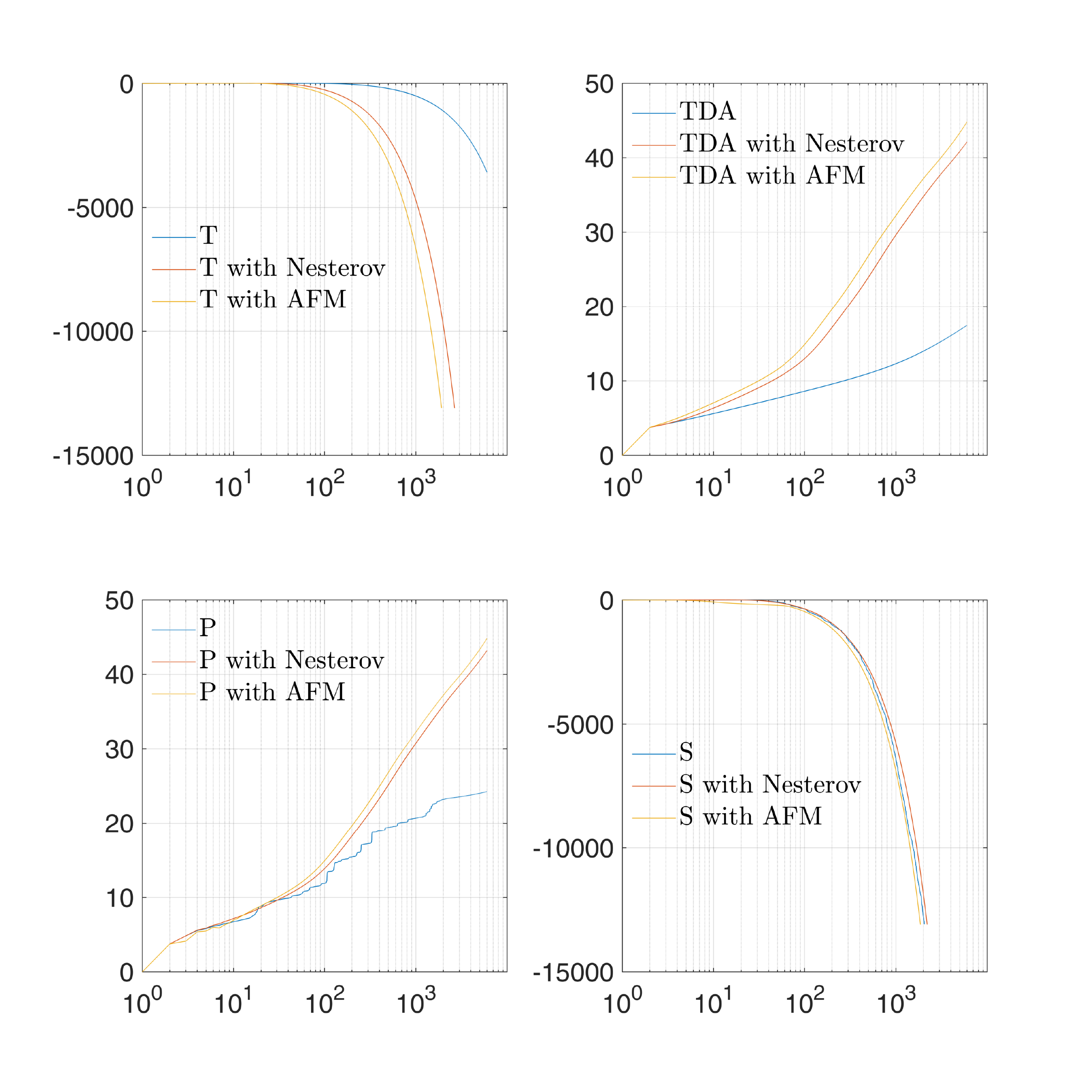}
         \caption{Gaussian ($\sigma=5$)}
     \end{subfigure}
\bigskip
 \begin{subfigure}[b]{0.49\textwidth}
         \centering
         \includegraphics[width=\textwidth,viewport=35bp 25bp 690bp 730bp,clip]{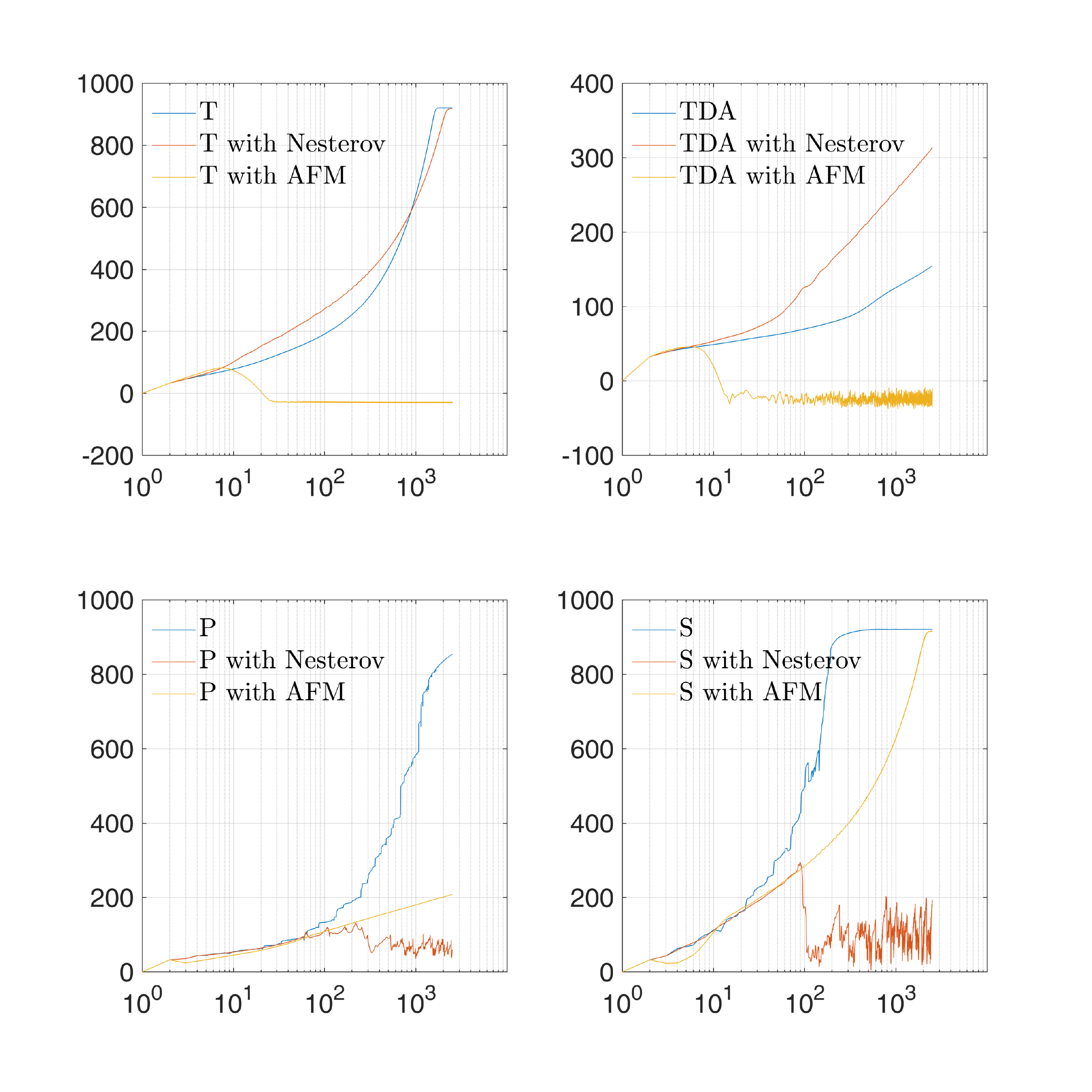}
         \caption{Guided}
     \end{subfigure}
\hspace{1mm}
     \begin{subfigure}[b]{0.49\textwidth}
         \centering
         \includegraphics[width=\textwidth,viewport=35bp 25bp 690bp 730bp,clip]{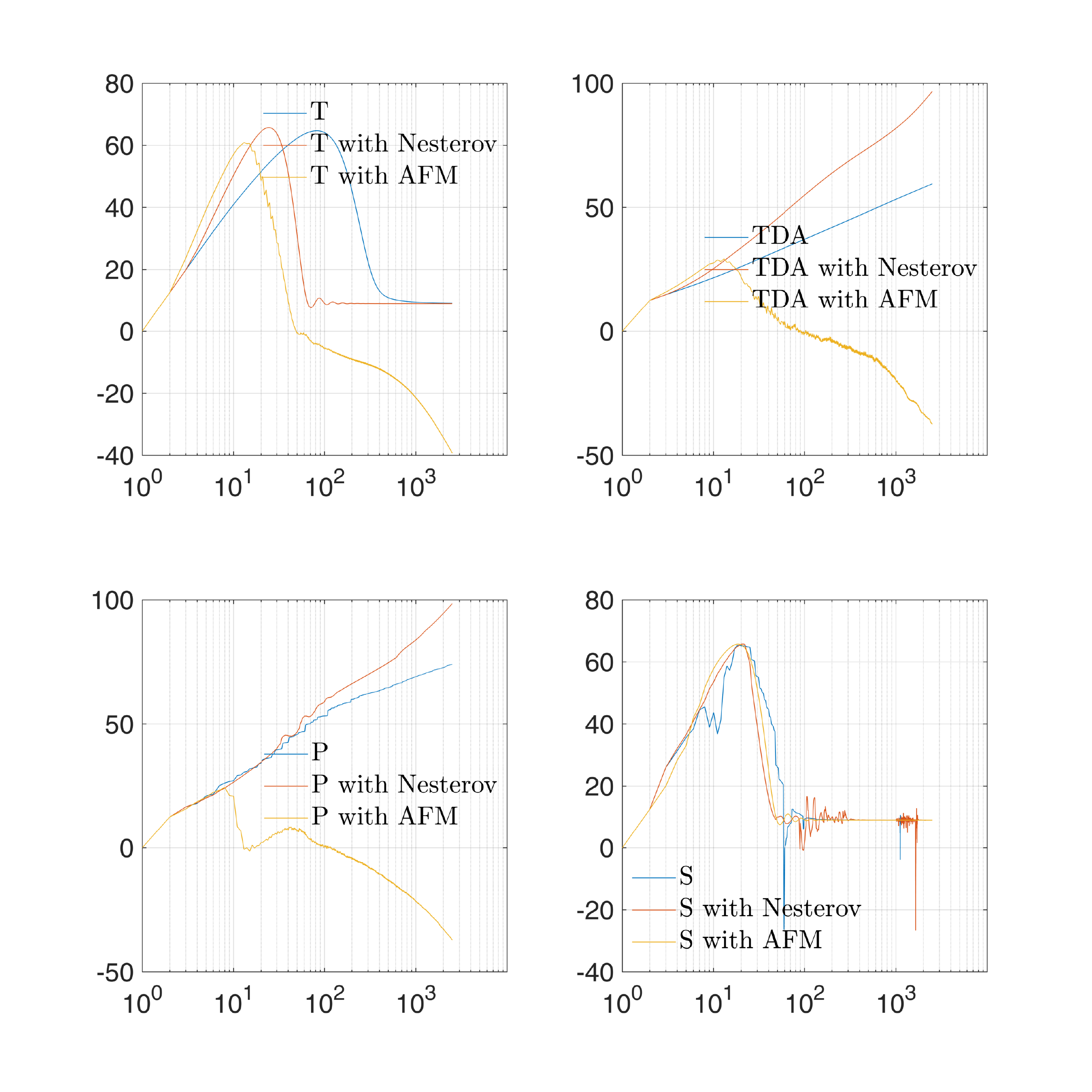}
         \caption{Bilateral}
     \end{subfigure}
        \caption{Results of the T, TDA, P and S methods.}
        \label{fig:three graphs}
\end{figure}

\clearpage
\begin{table}[t!]
\centering
   \begin{subtable}[h]{0.4\textwidth}
	\centering
    	\begin{tabular}{|c|c|c|c|c|}
    		\hline 
			& T & TDA & P & S \\
			\hline        			
			w/o Acc & 191 & 58 & 98 & 717 \\
			\hline    			
			with Nest. & 336 & 143 & - & - \\
			\hline          
     		with AFM & 350 & 148 & 148 & - \\  
     	   \hline                 
		\end{tabular}
    \caption{Gaussian filter $\sigma=1$.}
    \label{tab:week1}
	\end{subtable}
    \quad \quad
\begin{subtable}[h]{0.4\textwidth}
	\centering
       \begin{tabular}{|c|c|c|c|c|}     
            \hline 
            & T & TDA & P & S \\   
            \hline     
            w/o Acc & - & 17 & 24 & - \\    
            \hline         
            with Nest. & - & 42 & 43 & - \\        
            \hline    
            with AFM & - & 45 & 45 & - \\     
            \hline                  
        \end{tabular}
        \caption{Gaussian filter $\sigma=5$.}
        \label{tab:week2}
     \end{subtable}
     \medskip
	\begin{subtable}[h]{0.4\textwidth}
	\centering
		\begin{tabular}{|c|c|c|c|c|}                                                          
			\hline
			& T & TDA & P & S \\
			\hline       
			w/o Acc & 921 & 155 & 854 & 920 \\                                       \hline
			with Nest. & 918 & 313 & 60 & 187 \\                                     \hline
			with AFM & - & - & 208 & 916 \\   
			\hline  
		\end{tabular}                                                                         
		\caption{Guided filter.}
	\end{subtable}
\quad \quad
\begin{subtable}[h]{0.4\textwidth}
\centering
    \begin{tabular}{|c|c|c|c|c|}
        \hline   
        & T & TDA & P & S \\
		\hline
		w/o Acc & 9 & 59 & 74 & 9 \\
		\hline
		with Nest. & 9 & 97 & 98 & 9 \\
		\hline
		with AFM & - & - & - & 9 \\
		\hline 
    \end{tabular}                                                                          
\caption{Bilateral filter.}
\end{subtable}
\caption{Percentage of PSNR improvement at the end of iteration T/TDA/P/S methods. The entry with "-" means the method is divergent.}
\label{tab:subtables}
\end{table}

\end{document}